\documentclass[11pt,3p,review,authoryear]{elsarticle}

\journal{Expert Systems With Applications}
\biboptions{longnamesfirst}

\usepackage{natbib}
\usepackage{hyperref}
\usepackage{graphicx}   
\usepackage{float}
\usepackage{subcaption}
\usepackage{xcolor} 
\usepackage{balance}
\usepackage{amsmath}
\usepackage{amsfonts}
\usepackage{booktabs}
\usepackage{array}

\hypersetup{
    colorlinks=true,
    linkcolor= blue,
    citecolor= blue,
    urlcolor= blue
}

\begin{document}

\begin{frontmatter}

\title{Stockformer: A Price-Volume Factor Stock Selection Model Based on Wavelet Transform and Multi-Task Self-Attention Networks}


\author[ss]{Bohan Ma\fnref{note1}}
\author[ss]{Yushan Xue\fnref{note1}}
\author[ss]{Yuan Lu}
\author[ss]{Jing Chen\corref{cor1}}
\address[ss]{School of Statistics and Mathematics, Central University of Finance and Economics, Beijing 100081, China}

\cortext[cor1]{Corresponding author}
\ead{chenjingma@cufe.edu.cn}

\fntext[note1]{Bohan Ma and Yushan Xue contributed equally to this work.}

\begin{abstract}
As the Chinese stock market continues to evolve and its market structure grows increasingly complex, traditional quantitative trading methods are facing escalating challenges. Particularly, due to policy uncertainty and the frequent market fluctuations triggered by sudden economic events, existing models often struggle to accurately predict market dynamics. To address these challenges, this paper introduces ``Stockformer'', a price-volume factor stock selection model that integrates wavelet transformation and a multitask self-attention network, aimed at enhancing responsiveness and predictive accuracy regarding market instabilities. Through discrete wavelet transform, Stockformer decomposes stock returns into high and low frequencies, meticulously capturing long-term market trends and short-term fluctuations, including abrupt events. Moreover, the model incorporates a Dual-Frequency Spatiotemporal Encoder and graph embedding techniques to effectively capture complex temporal and spatial relationships among stocks. Employing a multitask learning strategy, it simultaneously predicts stock returns and directional trends. Experimental results show that Stockformer outperforms existing advanced methods on multiple real stock market datasets. In strategy backtesting, Stockformer consistently demonstrates exceptional stability and reliability across market conditions—whether rising, falling, or fluctuating—particularly maintaining high performance during downturns or volatile periods, indicating a high adaptability to market fluctuations. To foster innovation and collaboration in the financial analysis sector, the Stockformer model's code has been open-sourced and is available on the GitHub repository: \url{https://github.com/Eric991005/Multitask-Stockformer}.

\end{abstract}

\begin{keyword}
Price-Volume Factor Selection \sep Stockformer \sep Wavelet Transform \sep Spatiotemporal Graph Embedding \sep TopK-Dropout Strategy
\end{keyword}

\end{frontmatter}

\section{Introduction}

Forecasting stock returns has long been a widely researched topic in the field of finance. While the classical efficient market hypothesis posits that stock market prices, based on publicly available information, are unpredictable \citep{schwartz1970efficient}, more recent studies indicate that variables such as interest rates, inflation, and investor sentiment can significantly predict future stock market returns \citep{bollerslev2014stock, ang2007stock, campbell2008predicting}. Beyond market returns, the ability to predict cross-sectional individual stock returns—which has identified over 400 stock characteristics, humorously termed the "factor zoo"—raises questions about the extent to which individual stock returns can be predicted and which stock characteristics truly provide useful information for out-of-sample return prediction \citep{cochrane2011presidential, harvey2016and, green2017characteristics}. Exploring these questions within the Chinese capital market, which has an A-share market capitalization of RMB 68 trillion, is crucial for improving the effective allocation of these vast financial resources.

The challenge of predicting out-of-sample stock returns in China arises from several issues. First, the multitude of factors affecting stock returns and the low signal-to-noise ratio in high-dimensional sparse matrices make it difficult for traditional econometric models to extract meaningful information. Second, the functional relationships between predictive features and stock returns are uncertain \citep{campbell1999force, he2013intermediary}, posing challenges in capturing the nonlinear structures between them. Third, given its relatively short history of just over two decades and the continuous evolution of its regulatory framework, the Chinese stock market is particularly susceptible to policy shifts and other abrupt events that significantly impact the market \citep{li2020economic, wang2017policy}.

In recent years, deep learning has become an indispensable tool in quantitative investment, particularly in enhancing multifactor strategies that form the basis for understanding stock price movements \citep{wang2022intelligent}. By automating feature learning and capturing nonlinear relationships in stock market data, deep learning algorithms effectively identify complex patterns, thus enhancing prediction accuracy \citep{guo2022quant}. While the global research community recognizes the potential of deep neural networks, such as Recurrent Neural Networks (RNNs) and Convolutional Neural Networks (CNNs), for predicting stock and futures prices, preliminary studies have shown promising results \citep{senxin2023research, yue2022applications}. However, the use of deep learning models like RNNs and CNNs, although extensive, seldom explores deeper neural network models that mine and construct market and yield sequence information, suggesting room for further development in the application of deep learning to stock markets.

This paper proposes a multitask prediction model for stocks based on wavelet transformation and self-attention networks. The main contributions of this work are:

\begin{enumerate}
  \item \textbf{Multifactor Model Input:} We constructed 360 price-volume factors as the primary feature inputs for the Stockformer model. These factors, rigorously tested statistically, cover multiple dimensions such as price, volume, and volatility, ensuring comprehensive market information capture.
  \item \textbf{High and Low Frequency Decomposition:} Discrete wavelet transformation is used to decompose returns into high and low frequencies. This technique allows the model to capture both short-term market fluctuations (including abrupt events) and long-term trends, enhancing detail detection in market dynamics.
  \item \textbf{Spatiotemporal Encoder and Attention Mechanism:} A dual-channel spatiotemporal encoder integrated with a fusion attention mechanism precisely predicts the returns and trend movements of individual stocks. This method handles both temporal dependencies and spatial correlations among stocks, thereby improving prediction accuracy and efficiency.
  \item \textbf{Graph Embedding Techniques:} By constructing spatial relationship graphs and temporal graphs of stocks, graph embedding methods are employed to deeply analyze the spatial and temporal characteristics of stocks, aiding the model in capturing dynamic market relationships.
  \item \textbf{Multitask Learning Strategy:} The multitask learning approach not only predicts the returns of stocks but also their upward or downward trends. This multi-output framework enables the model to provide more comprehensive market change predictions, enhancing the robustness of the forecasts.
\end{enumerate}

\section{Related Work}
\subsection{Factor Construction}
A factor is a quantifiable variable that influences stock returns, encapsulating the underlying economic and financial dynamics. The conventional factors encompass market, style, industry, macroeconomic, price-volume, and machine learning-derived factors. The seminal works by Fama et al. \citep{fama1992cross, fama2015five} introduced market factors such as excess returns and volatility, encapsulating the overall market performance and risk. Green et al. \citep{green2017characteristics} utilized cross-sectional regression on 94 US stocks to investigate style factors, shedding light on the relative performance of diverse stock categories. More recent work by Fan et al. \citep{fan2016projected} explores time-varying factor loadings, enhancing the adaptability of factor models to changing market conditions. Furthermore, Jensen et al. \citep{quantpedia2021beta} discusses the stability of beta-adjusted equity factors, confirming the presence of seasonal and momentum effects in the cross-section of factor returns.

Our research focuses on utilizing the price-volume factor of Alpha360 to optimize the inputs of deep learning models and improve the prediction performance.The Alpha360 dataset includes daily data on six fundamental stocks, which is used to form comprehensive data inputs through historical backtracking and feature construction to enhance the model's ability to capture market dynamics.

\subsection{Stock Return Prediction Models}

Recent breakthroughs in machine learning in dimensionality reduction, penalty terms, and functional techniques naturally excel in addressing the challenges mentioned in the Introduction extracting effective information and dealing with non-linear relationships in stock market data. Recent papers have explored various types of machine learning algorithms for predicting stock returns. \textbf{The first category} includes commonly used dimensionality reduction models in finance. These models compress high-dimensional data into lower dimensions while preserving key information. For instance, Ampomah et al. \citep{ampomah2020evaluation} and Qolipour et al. \citep{qolipour2021predictability} employed Principal Component Analysis (PCA) to simplify sets of fundamental features and technical indicators, and integrated PCA with tree-based machine learning classifiers to predict stock returns and price movements. Moreover, Zhong et al. enhanced the accuracy of stock return predictions significantly by applying various dimensionality reduction techniques such as PCA, Fuzzy Robust PCA (FRPCA), and Kernel-based PCA (KPCA), demonstrating the effectiveness and potential of reduction techniques in handling complex financial datasets. \textbf{The second category} encompasses linear models with penalty terms, which excel by incorporating penalty terms to reduce noise information load, thus enhancing prediction accuracy. Such models have shown excellent performance in the financial domain. For example, Yacine et al. \citep{ait2022and} explored the predictability of high-frequency stock returns using methods like LASSO. Xuemei et al. \citep{xuemei2024group} proposed a multi-logistic regression model combined with penalty terms such as G-LASSO, G-SCAD, and G-MCP, effectively improving the predictive power for stock return direction in high-dimensional data settings. \textbf{The third category} includes nonlinear models, which can fit the nonlinear structure between predictive variables and returns based on historical data. Scholars using artificial intelligence algorithms like Random Forest, Fuzzy Neural Networks, and Long Short-Term Memory (LSTM) networks have tested the effects of technical and macroeconomic predictive factors on daily stock price returns' explained variance \citep{fischer2018deep, sirignano2016deep, bao2017deep, butaru2016risk}. With continuous advancements in computing technology, deep learning has been widely applied in quantitative trading, particularly in predicting stock returns, as shown by Chong et al. \citep{chong2017deep}, who utilized an Autoencoder (AE) to transform raw data for use in Deep Neural Networks (DNNs), predicting future returns for 38 stocks in the Korean market. Dami et al. \citep{dami2021predicting} combined AE and LSTM models to optimize stock return predictions for ten companies on the Tehran Stock Exchange. Additionally, Gunduz \citep{gunduz2021efficient} used a Variational Autoencoder (VAE) to predict the hourly direction for eight banks in the BIST 30 index, demonstrating accuracy comparable to models trained with un-reduced features. Due to these advantages, machine learning technologies have become a frontier application in the financial domain, particularly in predicting financial market movements, processing textual information, and improving trading strategies.

The Chinese stock market remains in a phase of continuous development and improvement, and economic policy uncertainty \citep{baker2016measuring} renders the predictability of stock returns challenging. Numerous scholars have integrated machine learning and deep learning technologies to address the prediction of expected returns in the Chinese stock market. Jiang et al.\citep{jiang2011predictable} investigated the predictability of stock returns for portfolios categorized by industry, size, price-to-book ratios, and ownership concentration; Li et al. \citep{li2017sentiment} found that deep learning outperformed traditional econometric models in predicting the CSI 300 index. Moreover, considering the dynamic changes in stock market styles (concept drift), Song et al. \citep{song2023follow} proposed a Contextual Information Shift Perception (CISP) method that incorporates dynamic parameters in stock predictions, effectively adapting to changes in the Chinese market style and thus providing more accurate forecasts. Although economic policy uncertainty and its resultant abrupt events are widely recognized as key factors influencing stock market volatility \citep{cai2022economic, wang2021covid, li2021dynamic}, current market models often fail to identify the specific timing and impact of these events. Furthermore, existing models frequently do not adjust their predictions promptly after abrupt events occur, leading to prediction biases due to excessive reliance on factor momentum effects \citep{ehsani2022factor}. This phenomenon not only weakens the models' responsiveness to rapid market changes but also limits their practical application in trading.

This research introduces a model that utilizes advanced signal processing techniques—wavelet transformation—to identify and isolate the signals of abrupt events within stock return sequences. Furthermore, by incorporating a self-attention mechanism, the model ensures rapid adaptation to market changes following such events, thereby enhancing the accuracy and timeliness of its predictions.

\subsection{Backtesting Trading Strategy}

Backtesting trading strategies is an integral part of the investment process, enabling the assessment of strategy performance and risk management before real-world application. Fung and Hsieh \citep{fung1997empirical} provide an empirical examination of hedge funds' dynamic trading strategies, establishing a framework for backtesting and evaluation. Georgakopoulos \citep{georgakopoulos2015quantitative} demonstrates the utility of automated programming for backtesting, detailing processes from data acquisition to strategy implementation and risk assessment. Chan \citep{chan2021quantitative} offers a thorough overview of quantitative trading foundations, with a focus on backtesting, data handling, and strategy execution. De Prado's work \citep{de2018advances} highlights the convergence of machine learning with financial backtesting, introducing "out-of-sample" testing to enhance strategy robustness. Qian \citep{qian2007quantitative} discusses quantitative equity portfolio management, covering backtesting and optimization across strategy development and performance evaluation.Together, these contributions provide theoretical frameworks, practical insights, and extensive case studies for backtesting trading strategies. Despite these advancements, there is still unexplored territory in backtesting, suggesting room for further research and innovation in strategy development and validation.

This study conducts backtesting using the Qlib framework developed by Microsoft\footnote{Qlib: \url{https://github.com/microsoft/qlib}}, employing the TopK-Dropout trading strategy, and compares it with traditional and advanced models to validate the effectiveness and stability of the Stockformer model. This approach not only enhances the transparency of the strategy but also ensures its stability across different market conditions.

\section{Preliminaries}

\subsection{Problem Definition}
The definition of stock prediction problems varies depending on the specific investment strategy adopted by investors. In this paper, we employ a widely accepted paradigm in the research field, the panel data analysis method \citep{muhammad2018relationship}. This approach integrates the returns, trend movements, and 360 price-volume factors of multiple stocks over a past period to predict the future returns and trend movements of multiple stocks for an upcoming period. This method fully utilizes multidimensional data resources and provides a theoretical foundation and practical guidance for building robust investment portfolios and exploring potential arbitrage opportunities by considering various financial factors, including both fundamental and technical aspects.

Consider the task of predicting the returns and trend movements of multiple stocks using the Stockformer method. Let $x_t^i \in \mathbb{R}^{362}$ represent the feature vector of stock characteristics for the $i$-th stock at time step $t$, consisting of the return rate (1 dimension), trend direction (1 dimension), and the 360 dimensions from the Alpha360 factor library, where $i \in \{1, \ldots, N\}$, and $N$ is the number of stocks considered. The feature tensor of all $N$ stocks at time step $t$ is denoted as $X_t = \left[x_t^1, \ldots, x_t^i, \ldots, x_t^N\right]^T \in \mathbb{R}^{N \times 362}$. Given a historical dataset of stock features: $\mathcal{X} = \left\{X_1, \ldots, X_{T_1}\right\} \in \mathbb{R}^{T_1 \times N \times 362}$, where $T_1$ is the number of historical time steps, the objective is to predict the returns (1 dimension) and trend movements (0 for a downtrend, 1 for an uptrend, 1 dimension) for the subsequent $T_2$ time steps. The predicted stock indicators are denoted as $\hat{\mathcal{Y}} = \left\{\hat{Y}_{T_1+1}, \ldots, \hat{Y}_{T_1+T_2}\right\} \in \mathbb{R}^{T_2 \times N \times 2}$, where $\hat{Y}_t \in \mathbb{R}^{N \times 2}$ represents the predicted stock indicators at time step $t$. The true values of these predictions are represented by $\mathcal{Y}_{\text{true}} = \left\{Y_{T_1+1}, \ldots, Y_{T_1+T_2}\right\} \in \mathbb{R}^{T_2 \times N \times 2}$.

The challenge lies in accurately forecasting $\hat{\mathcal{Y}}$ based on the patterns and trends identified in $\mathcal{X}$, while considering various factors that influence stock dynamics.

\subsection{Self-Attention Mechanism}
The self-attention mechanism\citep{vaswani2017attention} is the most commonly used attention mechanism that improves the accuracy of stock return forecasts by allowing the model to capture global dependencies across the entire input sequence. The mechanism processes inputs composed of queries, keys, and values, each having a dimension $d$. The procedure involves computing the dot products of the query with all the keys, dividing each by $\sqrt{d}$, and then applying a softmax function to obtain the weights on the values. Mathematically, this operation can be represented as:

\begin{equation}\label{Self-Attention}
\operatorname{Att}(Q, K, V)=\operatorname{softmax}\left(\frac{(Q W^Q)(K^T W^K)}{\sqrt{d}}\right)(V W^V),
\end{equation}
where $W^Q$, $W^K$, and $W^V$ are the learnable parameters of the projections, and $\operatorname{Att}(\cdot)$ denotes the self-attention operation.

\subsection{Wavelet Transform}
Wavelet transform, introduced by Daubechies \cite{daubechies1992ten}, is a mathematical tool used for the hierarchical decomposition of signals. It involves scaling functions and wavelet functions, which create a stable basis in the signal space through their shifts and expansions. This study explores the effectiveness of discrete wavelet transform in financial time series such as the return sequence of a single stock. For a one-dimensional stock return sequence \( r_t \in \mathbb{R}^T \), the discrete wavelet transform is applied to decompose the return sequence into low and high frequency components:
\begin{equation}
r_{l_k} = \sum_j g_{j-2k} r_j, \quad r_{h_k} = \sum_j h_{j-2k} r_j,
\end{equation}
where \( \mathbf{g} = \left\{g_k\right\}_{k \in \mathbb{Z}} \) and \( \mathbf{h} = \left\{h_k\right\}_{k \in \mathbb{Z}} \) are the low-pass and high-pass filters, respectively. The low-frequency component \( r_l \) captures the long-term trends of stock returns, while the high-frequency component \( r_h \) reflects short-term fluctuations and abrupt events. This decomposition aims to provide new perspectives and analytical tools for understanding and predicting market dynamics.

\section{Methodology}

\subsection{Overall Model Architecture}

  \begin{figure}[!htbp]
  	\centering
  	\includegraphics[width=1\textwidth]{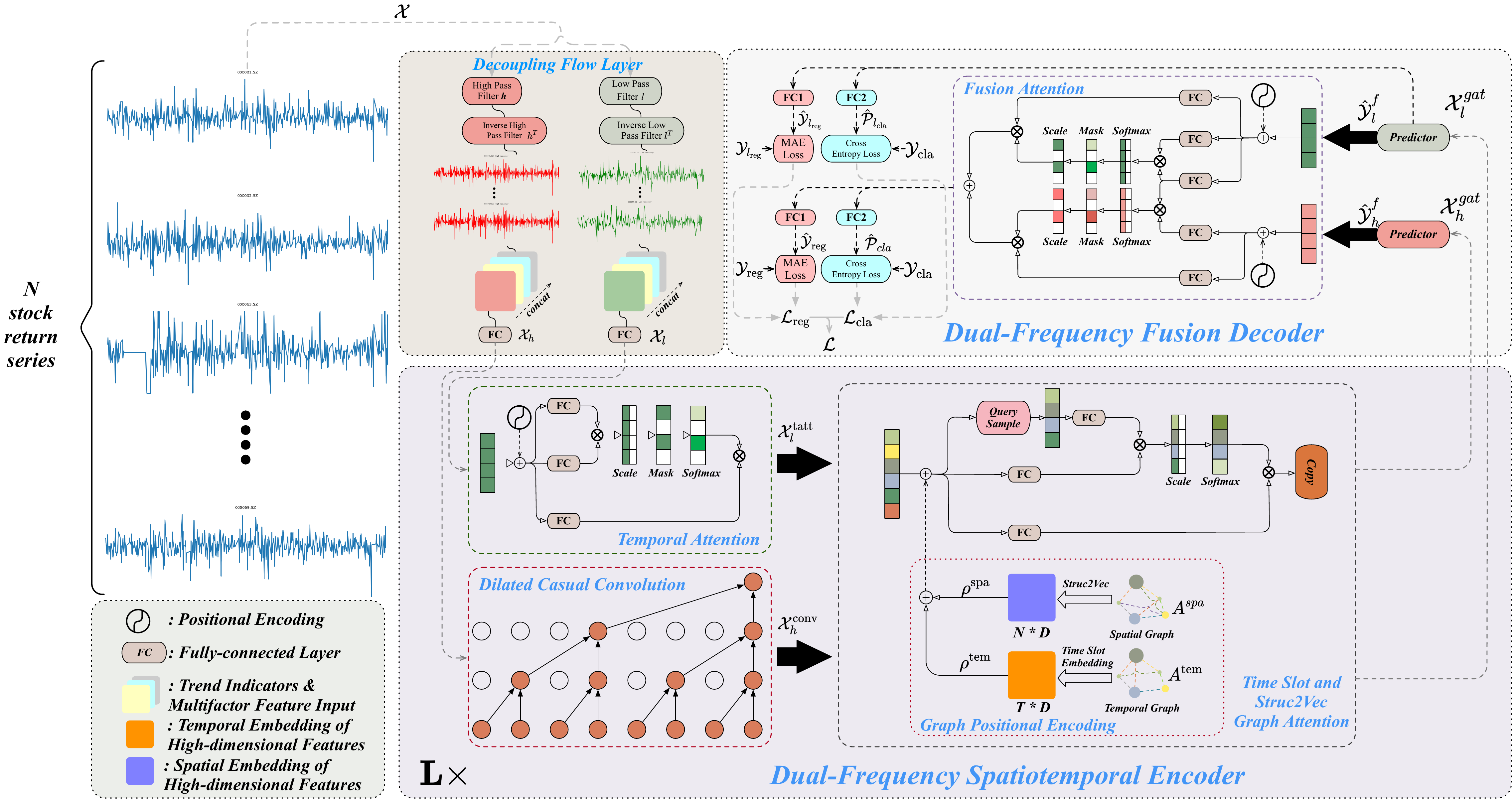} 
  	\caption{Stockformer architecture diagram, which primarily consists of three parts: the Decoupling Flow Layer, the Dual-Frequency Spatiotemporal Encoder, and the Dual-Frequency Fusion Decoder.} 
  	\label{fig:Multi-task_Stockformer} 
\end{figure}

Based on Figure \ref{fig:Multi-task_Stockformer}, the Stockformer architecture is primarily divided into three parts: the \textbf{Decoupling Flow Layer}, the \textbf{Dual-Frequency Spatiotemporal Encoder}, and the \textbf{Dual-Frequency Fusion Decoder}. Initially, the historical stock data $\mathcal{X} \in \mathbb{R}^{T_1 \times N \times 362}$ is processed through the \textbf{Decoupling Flow Layer}, where the stock return series tensor undergoes wavelet transformation to separate into \textbf{high} and \textbf{low frequency components}, while other parts (such as trend indicators and price-volume factors) remain unchanged. These components are then concatenated with the unchanged parts along the third dimension. Here, the \textbf{low-frequency component} (denoted as $l$) captures long-term trends, and the \textbf{high-frequency component} (denoted as $h$) captures short-term fluctuations and abrupt events. These are denoted as $\mathcal{X}_h, \mathcal{X}_l \in \mathbb{R}^{T_1 \times N \times 362}$, where $\mathcal{X}_h, \mathcal{X}_l$ are linearly transformed through a fully connected layer to $\mathbb{R}^{T_1 \times N \times D}$.

Subsequently, a \textbf{Dual-Frequency Spatiotemporal Encoder} is designed to represent these distinct time series patterns: the low-frequency features are fed into a \textbf{Temporal Attention} (denoted as $tatt$) module, while the high-frequency features are processed through a \textbf{Dilated Causal Convolutional Layer} (denoted as $conv$), represented as $\mathcal{X}_l^{\text{tatt}}, \mathcal{X}_h^{\text{conv}} \in \mathbb{R}^{T_1 \times N \times D}$. These components are then input into \textbf{Graph Attention Networks} (denoted as $gat$), interacting with graph information to enable the model to capture complex relationships and dependencies among stocks and time. In this module, the spatial graph $A^{spa}$ and temporal graph $A^{tem}$ are transformed through a fully connected layer and tensor broadcasting operations to high-dimensional embeddings denoted as $\rho^{\mathrm{spa}}, \rho^{\mathrm{tem}} \in \mathbb{R}^{T_1 \times N \times D}$, which are then fused with $\mathcal{X}_l^{tatt}, \mathcal{X}_h^{conv}$ through addition and undergo \textbf{graph attention operations} to produce $\mathcal{X}_l^{gat}, \mathcal{X}_h^{gat} \in \mathbb{R}^{T_1 \times N \times D}$. The \textbf{Dual-Frequency Spatiotemporal Encoder} consists of $L$ stacked layers, aimed at effectively representing the dual-scale spatiotemporal patterns of low and high-frequency waves. Finally, in the \textbf{Dual-Frequency Fusion Decoder}, predictors generate $\hat{\mathcal{Y}}_i^f, \hat{\mathcal{Y}}_h^f \in \mathbb{R}^{T_2 \times N \times D}$, which are aggregated through \textbf{Fusion Attention} interactions to obtain a latent representation of dual-scale temporal patterns $\hat{\mathcal{Y}}^f \in \mathbb{R}^{T_2 \times N \times D}$. Through distinct fully connected layers (regression layer FC1 and classification layer FC2), \textbf{multi-task outputs} are produced, including stock return predictions (regression result, denoted as $reg$) $\hat{\mathcal{Y}}_{reg} \in \mathbb{R}^{T_2 \times N}$ and stock trend prediction probabilities (classification result, denoted as $cla$) $\hat{\mathcal{P}}_{cla} \in \mathbb{R}^{T_2 \times N}$. Additionally, regression values for the low-frequency component $\hat{\mathcal{Y}}_{l_{reg}} \in \mathbb{R}^{T_2 \times N}$ and trend prediction probabilities $\hat{\mathcal{P}}_{l_{reg}} \in \mathbb{R}^{T_2 \times N}$ are output to enhance learning of low-frequency signals in the supervision signal, with detailed loss function calculations presented in the Multi-Supervision subsection \ref{subsubsec:multi_supervision}.

\subsection{Decoupling Flow Layer}
Given a historical stock data set $\mathcal{X}$, focusing on the stock return series, this study employs Discrete Wavelet Transform (DWT) from the Decoupling Flow Layer to extract the low and high frequency component sequences of returns, representing long-term and short-term temporal patterns respectively. The low-frequency component, stable and indicative of long-term trends, reflects the trajectory of returns; while the volatile high-frequency component reflects short-term fluctuations, such as those induced by economic policy uncertainties leading to abrupt market events. The DWT of the input return series $\mathcal{X}$ can be represented by Equation \ref{eq:DWT}:
\begin{equation} \label{eq:DWT}
\overline{\mathcal{X}}_l=\mathbf{g} \mathcal{X}, \quad \overline{\mathcal{X}}_h=\mathbf{h} \mathcal{X},
\end{equation}
where the candidate low and high frequency components $\overline{\mathcal{X}}_l$ and $\overline{\mathcal{X}}_h$ undergo a downsampling operation that reduces the time step length to half in DWT. Consequently, this layer uses the inverse low-pass and high-pass filters $\mathbf{g}^T$ and $\mathbf{h}^T$ to upsample the inputs. The trend indicators $(0,1)$ and the 360-dimensional multifactor features (Alpha360) are then concatenated with the upsampled low and high frequency components, forming the final multi-dimensional low and high frequency inputs. These are then transformed by a fully connected layer into high-dimensional low and high frequency components $\mathcal{X}_l, \mathcal{X}_h$, enhancing the representational power of the Stockformer. The upsampling operation and fully connected layer are formalized in Equation \ref{eq:upsample}:
\begin{equation} \label{eq:upsample}
\mathcal{X}_l=W^g \mathbf{g}^T \overline{\mathcal{X}}_l+b^g, \quad \mathcal{X}_h=W^h \mathbf{h}^T \overline{\mathcal{X}}_h+b^h,
\end{equation}
where $W^g, W^h \in \mathbb{R}^{C \times d}$ and $b^g, b^h \in \mathbb{R}^d$ are learnable parameters.

\subsection{Dual-Frequency Spatiotemporal Encoder}
The Dual-Frequency Spatiotemporal Encoder is comprised of three major components: Temporal Attention, Dilated Causal Convolution, and Time Slot with Struc2Vec Graph Attention Networks. 

\subsubsection{Decoupling Temporal Feature Extraction}
Unlike previous works that solely relied on methods like LSTM for processing financial sequences, our approach integrates time convolution layers and time attention to focus on trends and seasonal components. Time attention captures long-term low-frequency trends by considering global sequence relationships, while the time convolution layer focuses on local patterns, effectively simulating high-frequency components and abrupt events. This dual modeling approach enhances the prediction accuracy of complex financial sequences. Dilated causal convolution is a specific type of one-dimensional convolution that slides over the input by skipping values at a defined step, as shown in Figure \ref{fig:Multi-task_Stockformer}. Theoretically, given a one-dimensional sequence input $x \in \mathbb{R}^T$ and a filter $f \in \mathbb{R}^J$, the dilated causal convolution operation at time step $t$ is defined as:
\begin{equation} \label{eq:dc_convolution}
x \star f(t) = \sum_{j=0}^J f(j) x(t - c \times j),
\end{equation}
where $c$ is the dilation factor. The dilated causal convolution of the high-frequency component is expressed as:
\begin{equation} \label{eq:hf_dilated_convolution}
\mathcal{X}_h^{\text{conv}} = \operatorname{ReLU}(\Theta \star \mathcal{X}_h + b)
\end{equation}
where $\Theta$ and $b$ are learnable parameters, and $\operatorname{ReLU}(\cdot)$ is the rectified linear unit.

Here, self-attention is utilized to capture the long-term trends in the low-frequency component, reflecting the stability and prominent long-term trends observed in return trajectories:
\begin{equation} \label{eq:self_attention_tatt}
\begin{gathered}
\mathcal{X}_l^{\text{tatt}} = \operatorname{Concat}(ta_1, \ldots, ta_n, \ldots, ta_N), \\
\text{where } ta_n = \operatorname{Att}(X_l^n, X_l^n, X_l^n)
\end{gathered}
\end{equation}

\subsubsection{Time Slots and Struc2Vec Graph Attention Networks}

\paragraph{\textbf{Time Slots}}
As previously discussed, the stock returns exhibit cyclical fluctuations, with each day in the dataset represented by a timestamp $t$, which is considered as a unit of time. Thus, it is essential to extract time features from $t$. A straightforward approach is to treat each timestamp as a floating-point number and then use a multilayer perceptron model to convert this value into a fixed-length vector. However, this method has two disadvantages. Firstly, each timestamp is typically a large integer, which could dominate other features when used directly. Secondly, there are monthly and daily cyclic variations between different timestamps that cannot be captured merely by the timestamps. For example, trading conditions during the same period across different months might be similar. To address these issues, we employ a method of time slots to represent timestamps.

Given a baseline timestamp $t_0$ (to ensure $t-t_0 \geq 0$, $t_0$ must be less than any timestamp in the training and testing data), and a time unit $\Delta t$, we can establish intervals such as
\[
\begin{aligned}
\left[t_0, t_0+\Delta t\right),&\\
\left[t_0+\Delta t, t_0+2\Delta t\right),&\\
\cdots&
\end{aligned}
\]
These intervals are referred to as time slots, each with a size of $\Delta t$, which represents one day in this dataset.

For simplicity, each time slot is represented by its index number. For instance, $\left[t_0, t_0+\Delta t\right)$ is represented as 0. A timestamp $t$ can be mapped to a specific time slot $t^p$, where $t \geq t_0$, computed as shown in Equation \ref{tp}:
\begin{equation}\label{tp}
t^p=\left\lfloor\frac{t-t_0}{\Delta t}\right\rfloor
\end{equation}
For a finer granularity of each timestamp, the remainder $t^r$ is recorded to ensure the uniqueness of $t$, where $0 \leq t^r < \Delta t$, calculated as shown in Equation \ref{tr}:
\begin{equation}\label{tr}
t^r=t-t_0-t^p \Delta t
\end{equation}
In summary, each timestamp $t$ can be represented as $\left\langle t^p, t^r\right\rangle$.

\paragraph{\textbf{Constructing the Temporal Graph}}

Next, we explore how to capture temporal features by embedding time slots. Given the nature of the dataset, each time slot can represent a day. Considering the periodic fluctuations in stock returns and the monthly cycle being the most common cyclical fluctuation in economic cycles, it suffices to focus on all time slots within a month. With approximately 250 trading days in a year distributed across months, each month averages 21 trading days. Inspired by \cite{li2018multi}, we attempt to construct a temporal graph to represent time slots and then apply graph embedding methods to initialize the time slot embeddings. However, \cite{li2018multi} constructed an undirected graph for time slots, failing to capture the sequential relationship between them. Moreover, they overlooked the connection between adjacent dates, thus missing daily periodicity. 

\begin{figure}[!htbp]
  	\centering
  	\includegraphics[width=0.5\columnwidth]{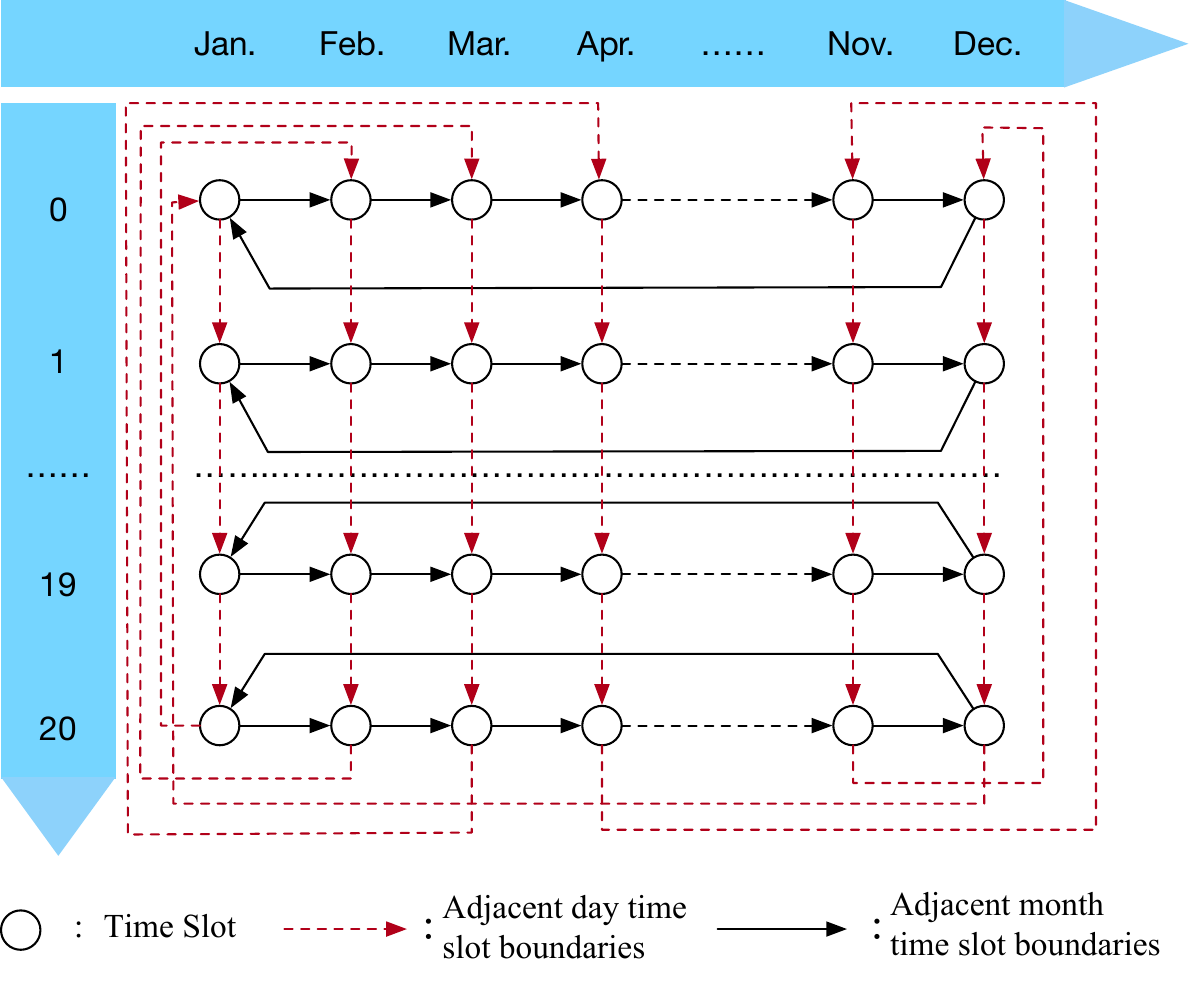} 
\caption{Temporal graph depicting a year with each time slot representing a day. Each red directed line connects two adjacent time slots, while each black directed line connects the same time slot in two adjacent months.}
  	\label{fig:Temporal_graph} 
  \end{figure}

Addressing the issue at hand, we adopt the time-slot method to represent time stamps, drawing inspiration from \cite{yuan2020effective}, as shown in Figure \ref{fig:Temporal_graph}.
 This representation is captured by the graph \(G^{\prime}=\left\langle V^{\prime}, E^{\prime}\right\rangle\). In this context, each node \(v^{\prime} \in V^{\prime}\) denotes a time slot. The edges in \(E^{\prime}\) are categorized into two types: (1) edges connecting adjacent date slots, symbolizing contiguous time slots, thus providing smoothness; and (2) edges connecting adjacent months, denoting identical time slots in consecutive months, ensuring similarity. As an illustration, this study designates \(\Delta t\) as 1 day, partitioning each month into 21 distinct time slots. Taking into account the twelve months within a year, we construct a directed temporal graph with a size of \(21 \times 12=252\). Ultimately, \(t^p\) can be mapped to a node \(v^{\prime} \in V^{\prime}\), where the sequence number for \(v^{\prime}\) is given by \(t^{p}\% 252\), with \(\%\) representing the modulus operator.

\paragraph{\textbf{Embedding Time Slots in the Temporal Graph}}

Initially, one-hot encoding is used to represent each time slot \(O_i^t \in \mathbb{R}^{\left|V^{\prime}\right|}\), where \(\left|V^{\prime}\right|\) is the total number of nodes in the temporal graph \(G^{\prime}\). Then, a fully connected neural network (with weight matrix \({W}*t \in \mathbb{R}^{\left|V^{\prime}\right| \times d_t}\)) is designed to transform each one-hot encoded \(O_i^t\) into a fixed-length dense vector \(D_i^t={W}*{{t}}^{\top} O_i^t\). This embedding is used as the initial value for \({W}_{{t}}\), resulting in a high-dimensional temporal graph embedding \(\rho^{\text {tem }}\in \mathbb{R}^{T_1 \times D} \). Leveraging the broadcasting property, \(\rho^{\text {tem }}\) is replicated N times, ultimately yielding \(\rho^{\text {tem }}\) with dimensions \(\rho^{\text {tem }}\in \mathbb{R}^{T_1 \times N \times D}\).

\paragraph{\textbf{Struc2Vec Graph Embedding}}
By computing the Spearman correlation coefficients between the returns data of $N$ stocks, we construct an $N \times N$ correlation matrix to serve as the adjacency matrix, which describes the connectivity between stocks. Each node represents a stock, and the weight of the edges reflects the correlation between stocks. In the Struc2Vec algorithm, each stock node is initially randomly initialized with a vector. Through multiple iterations, each node's vector representation is refined by integrating information from neighboring nodes and the weighted neighbor vectors using an attention mechanism. This process iterates until the vector representations stabilize or the predetermined number of iterations is reached. Ultimately, after multiple rounds of updates, each node's (stock's) vector representation will contain structural information about its position in the graph and the influence of its neighbors, thereby achieving high-dimensional stock relational graph embeddings $\rho^{spa} \in \mathbb{R}^{N \times d}$. Leveraging the property of broadcasting, $\rho^{spa}$ is replicated to match $T_1$ times, resulting in dimensions $\rho^{spa} \in \mathbb{R}^{T_1 \times N \times D}$.

\paragraph{\textbf{Self-Attention Encoding}}

After obtaining the high dimensional temporal graph embedding and the stock return association graph embedding, this study broadcasts and sums the two, concatenating them with the seasonal and trend components, respectively. This results in an input tensor for self-attention encoding. The final time slot and Struc2Vec graph attention network can be represented as in Equation \ref{attentionGraph}:
\begin{equation}\label{attentionGraph}
\begin{aligned}
& \mathcal{X}^{\text {gat }}=\operatorname{Concat}\left(s a_1, \ldots s a_t, \ldots, s a_{T_1}\right) \\
& \text { where } \quad \text { sa } a_t=\operatorname{Att}\left(\tilde{X}_t, \tilde{X}_t, \tilde{X}_t\right) \\
& \text { and } \tilde{\mathcal{X}}=\mathcal{X}+\rho^{\mathrm{spa}}+\rho^{\mathrm{tem}}
\end{aligned}
\end{equation}

Thus, through the time slot and Struc2Vec graph attention network mechanism, this study obtains \(\mathcal{X}^{\text {gat }}\)(The high-frequency signals are denoted as $\mathcal{X}_h^{\mathrm{gat}}$ and the low-frequency signals as $\mathcal{X}_l^{\mathrm{gat}}$, with both types undergoing the same processing steps here. ), a vector containing structural information and localized graph characteristics. This attention mechanism is highly efficient and expressive when handling graph data.

\subsection{Dual-Frequency Fusion Decoder}

In this study, predictors (i.e., fully connected layers) are applied on the temporal dimensions of the representations encoded by the Dual-Frequency Spatiotemporal Encoder ($\mathcal{X}_l^{\text{gat}}$ and $\mathcal{X}_h^{\text{gat}}$) to transform them into future representations of multi-step stock returns and trends, resulting in future representations of low-frequency and high-frequency components $\hat{\mathcal{Y}}_i^f$ and $\hat{\mathcal{Y}}_h^f$. Given that high-frequency components often correspond to rapid changes or noise in the data, which are challenging to predict and uncertain, and low-frequency components generally correspond to long-term trends and global patterns, which are more persistent and stable, crucial for forecasting overall trends and long-term changes. Therefore, through fusion attention and a multi-supervision strategy, this paper integrates the information of low and high-frequency components, with particular emphasis on supervising the low-frequency components to extract useful long-term trend information.

\subsubsection{Decoupling Feature Fusion}
As the aim of this study is not to predict low and high-frequency components but to forecast future stock return sequences and trends based on these components and price-volume factors, as shown in Figure \ref{fig:Multi-task_Stockformer}, we further propose a fusion attention mechanism. This mechanism integrates the representations of low-frequency and high-frequency components $\hat{\mathcal{Y}}_l^f, \hat{\mathcal{Y}}_h^f$ into the stock returns $\hat{\mathcal{Y}}^f$ and captures future internal dependencies. Specifically, the fusion attention considers the low-frequency component as the query, extracting useful long-term and short-term information from both low and high-frequency components in two time attention mechanisms. The fusion attention can be expressed as in Equation \ref{eq:fusion_attention}:
\begin{equation} \label{eq:fusion_attention}
\begin{aligned}
& \hat{\mathcal{Y}}^f = \operatorname{Concat}\left(f a_1, \ldots, f a_n, \ldots, f a_N\right) \\
& \text{where} \quad f a_n = \operatorname{Att}\left(\hat{\mathcal{Y}}_l^{f^n}, \hat{\mathcal{Y}}_l^{f^n}, \hat{\mathcal{Y}}_l^{f^n}\right) + \operatorname{Att}\left(\hat{\mathcal{Y}}_l^{f^n}, \hat{\mathcal{Y}}_h^{f^n}, \hat{\mathcal{Y}}_h^{f^n}\right).
\end{aligned}
\end{equation}

In Equation \ref{eq:fusion_attention}, the study concatenates each stock $\mathrm{n}$'s attention mechanism output $f a_n$ to obtain $\hat{\mathcal{Y}}^f$. The attention mechanism calculates self-attention for low-frequency components and attention between low-frequency and high-frequency components. This fusion attention mechanism allows simultaneous utilization of both low and high-frequency component information, thereby better capturing future stock returns and internal dependencies.

\subsubsection{Multi-Supervision}
\label{subsubsec:multi_supervision}

During the training process, through Stockformer, the model is capable of making multi-dimensional predictions on future trends of stock sequences. Specifically, the model's output layer is divided into classification and regression task outputs, respectively providing the probabilities of stock trend predictions $\hat{\mathcal{P}}_{cla}$, low-frequency component trend predictions $\hat{\mathcal{P}}_{l_{cla}}$, predicted stock return values $\hat{\mathcal{Y}}_{\text{reg}}$, and low-frequency component values $\hat{\mathcal{Y}}_{l_{\text{reg}}}$.

\paragraph{\textbf{Loss Function for Multi-Supervision Optimization}}
As depicted in the Dual-Frequency Fusion Decoder in Figure \ref{fig:Multi-task_Stockformer}, the training integrates the loss functions of regression and classification tasks, allowing the model to learn multiple types of outputs simultaneously. The loss function is defined in Equation \ref{eq:loss_function}:
\begin{equation} \label{eq:loss_function}
\mathcal{L} = \mathcal{L}_{\text{reg}} + \lambda \mathcal{L}_{\text{cla}}
\end{equation}
where $\lambda$ is a hyperparameter used to balance the weights of regression and classification losses.

The computations for $\mathcal{L}_{\text{reg}}$ and $\mathcal{L}_{\text{cla}}$ are as follows:

\textbf{(1) Regression Loss $\mathcal{L}_{\text{reg}}$:} Used to measure the accuracy of the model in predicting the returns and their low-frequency components, calculated using the Mean Absolute Error (MAE) loss. Defined as shown in Equation \ref{eq:reg_loss}:
\begin{equation} \label{eq:reg_loss}
\mathcal{L}_{\text{reg}} = \frac{1}{N(T_2 - T_1)} \sum_{t=T_1+1}^{T_1+T_2} \sum_{n=1}^N \left( \left| y_t^n - \hat{y}_t^n \right| + \left| y_{l_t}^n - \hat{y}_{l_t}^n \right| \right)
\end{equation}
where $N$ is the number of samples, $(T_2-T_1)$ is the number of time steps, $y_t^n$ is the actual return of sample $n$ at time $t$, $\hat{y}_t^n$ and $\hat{y}_{l_t}^n$ are the predicted return and low-frequency component values, respectively.

\textbf{(2) Classification Loss $\mathcal{L}_{\text{cla}}$:} Used to assess the performance of the model in predicting stock trends, typically using a cross-entropy loss function. Defined as shown in Equation \ref{eq:cla_loss}:
\begin{equation} \label{eq:cla_loss}
\mathcal{L}_{\text{cla}} = \frac{1}{N(T_2-T_1)} \sum_{t=T_1+1}^{T_2} \sum_{n=1}^N \left(-\sum_k y_{c, t, k}^n \log \left(\hat{p}_{cla, t, k}^n\right) - \sum_k y_{c, t, k}^n \log \left(\hat{p}_{l_{cla}, t, k}^n\right)\right)
\end{equation}

where $y_{c, t, k}^n$ is the true class label for sample $n$ at time $t$ for class $k$. $\hat{p}_{cla, t, k}^n$ and $\hat{p}_{l_{cla}, t, k}^n$ are the predicted probabilities for the regular and low-frequency classification tasks, respectively. By minimizing the above loss functions, the model can optimize the prediction of stock trends and returns simultaneously, thereby enhancing the overall predictive performance and robustness of the model.


\section{Construction of Price-Volume Factors}
\label{sec:factor_construction}
\subsection{Data Sources and Stock Pool Selection}
\label{subsec:data_sources_stock_selection}
The CSI 300 Index and the China Securities Index series are currently the mainstream indices in the A-share market. The CSI 300 Index, consisting of the top 300 stocks in terms of liquidity and market capitalization from both Shanghai and Shenzhen stock exchanges, effectively reflects the overall trend of the A-share market and has good market representativeness. Therefore, this paper selects the constituent stocks of the CSI 300 Index as the stock pool for constructing the stock selection model.

\begin{table}[h]
\centering
\caption{Stock Market Indicators}
\label{tab:market_indicators}
\begin{tabular}{@{}ccccccc@{}}
\toprule
\textbf{Market Indicator} & \textbf{Open} & \textbf{High} & \textbf{Low} & \textbf{Close} & \textbf{VWAP} & \textbf{Volume} \\ 
\midrule
\textbf{Indicator Name}   & open          & high          & low           & close         & vwap          & volume         \\ 
\bottomrule
\end{tabular}
\end{table}

This study analyzes the stock market data from March 1, 2018, to January 30, 2024. Considering potential adjustments such as dividends and stock splits, this paper applies forward adjustments to the price-related indicators. Market indicators are presented in Table \ref{tab:market_indicators}.

\begin{figure}[h]
\centering
\includegraphics[width=\textwidth]{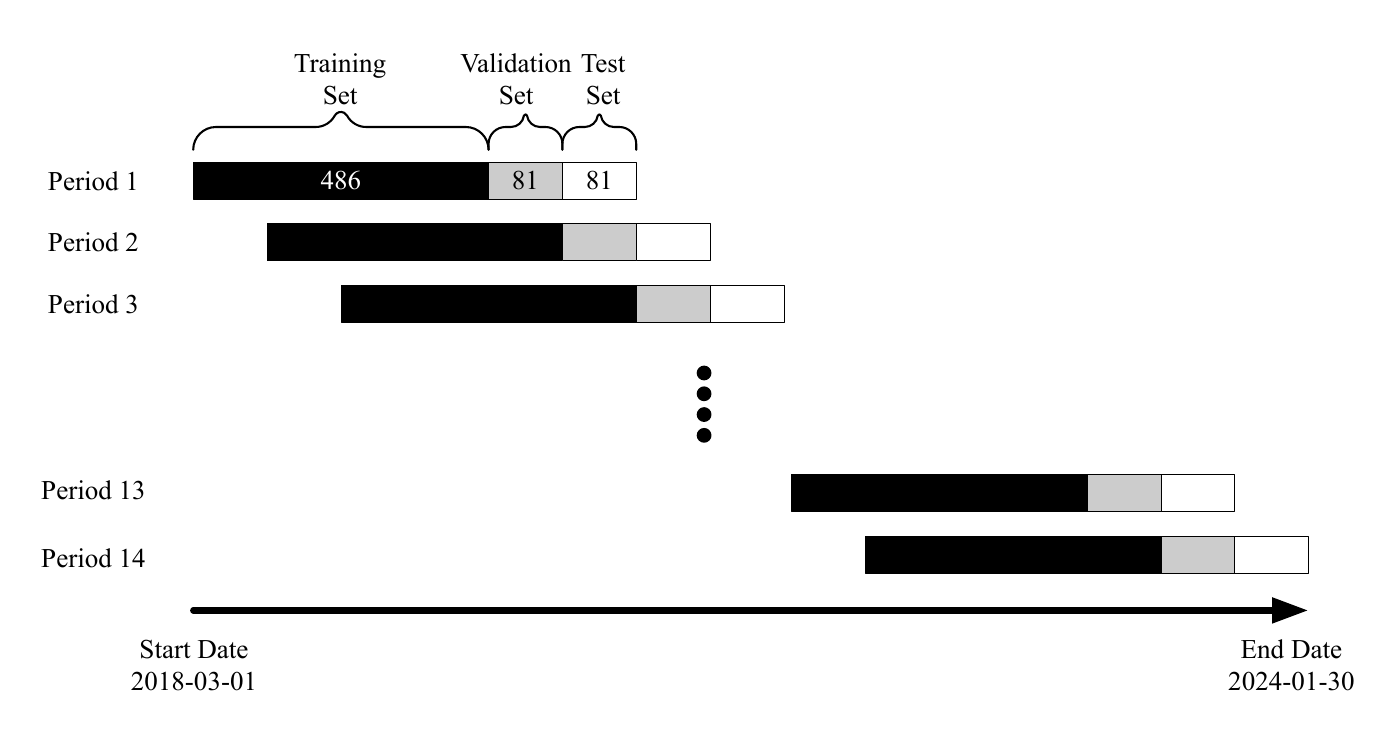}
\caption{Visualization of the dataset division method used in this study. Each segment represents a specific period in the rolling window analysis, showing the division into training, validation, and test sets across the specified dates.}
\label{fig:dataset_division}
\end{figure}

We utilize stock market data from March 1, 2018, to January 30, 2024, for analysis. As shown in Figure \ref{fig:dataset_division}, to maintain the model's generalizability and adaptability over different periods, a rolling window approach is used to generate sub-datasets. Considering the Chinese stock market is open for about 243 days annually, the data is divided into two years of trading data as the training set (486 days), followed by three months of trading data for the validation set (81 days), and another three months for the test set (81 days), resulting in 14 sub-datasets. These are then split by time into training (75\%), validation (12.5\%), and testing sets (12.5\%) as detailed in Table \ref{tab:detailed_subdataset_division}.

\begin{table}[h]
\centering
\caption{Subdataset Division with Detailed Date Ranges}
\label{tab:detailed_subdataset_division}
\resizebox{\textwidth}{!}{ 
\begin{tabular}{c|ccc|ccc|ccc}
\hline
 & \multicolumn{3}{c|}{\textbf{Training Set (486 days)}} & \multicolumn{3}{c|}{\textbf{Validation Set (81 days)}} & \multicolumn{3}{c}{\textbf{Test Set (81 days)}} \\
\cline{2-10} 
\textbf{Dataset} & \textbf{Start Date} & \textbf{End Date} & \textbf{Days} & \textbf{Start Date} & \textbf{End Date} & \textbf{Days} & \textbf{Start Date} & \textbf{End Date} & \textbf{Days} \\
\hline
Subdataset 1  & 2018-03-01 & 2020-02-28 & 486 & 2020-03-02 & 2020-06-30 & 81 & 2020-07-01 & 2020-10-29 & 81 \\
Subdataset 2  & 2018-05-31 & 2020-05-29 & 486 & 2020-06-01 & 2020-09-23 & 81 & 2020-09-24 & 2021-01-25 & 81 \\
Subdataset 3  & 2018-08-27 & 2020-08-26 & 486 & 2020-08-27 & 2020-12-25 & 81 & 2020-12-28 & 2021-04-28 & 81 \\
Subdataset 4  & 2018-11-28 & 2020-11-27 & 486 & 2020-11-30 & 2021-03-30 & 81 & 2021-03-31 & 2021-07-28 & 81 \\
Subdataset 5  & 2019-03-04 & 2021-03-02 & 486 & 2021-03-03 & 2021-06-30 & 81 & 2021-07-01 & 2021-11-01 & 81 \\
Subdataset 6  & 2019-06-03 & 2021-06-01 & 486 & 2021-06-02 & 2021-09-27 & 81 & 2021-09-28 & 2022-01-26 & 81 \\
Subdataset 7  & 2019-08-28 & 2021-08-26 & 486 & 2021-08-27 & 2021-12-28 & 81 & 2021-12-29 & 2022-05-05 & 81 \\
Subdataset 8  & 2019-11-29 & 2021-11-30 & 486 & 2021-12-01 & 2022-03-31 & 81 & 2022-04-01 & 2022-08-01 & 81 \\
Subdataset 9  & 2020-03-04 & 2022-03-03 & 486 & 2022-03-04 & 2022-07-04 & 81 & 2022-07-05 & 2022-11-02 & 81 \\
Subdataset 10 & 2020-06-03 & 2022-06-06 & 486 & 2022-06-07 & 2022-09-28 & 81 & 2022-09-29 & 2023-02-03 & 81 \\
Subdataset 11 & 2020-08-31 & 2022-08-30 & 486 & 2022-08-31 & 2022-12-29 & 81 & 2022-12-30 & 2023-05-05 & 81 \\
Subdataset 12 & 2020-12-02 & 2022-12-01 & 486 & 2022-12-02 & 2023-04-03 & 81 & 2023-04-04 & 2023-08-02 & 81 \\
Subdataset 13 & 2021-03-05 & 2023-03-06 & 486 & 2023-03-07 & 2023-07-05 & 81 & 2023-07-06 & 2023-11-03 & 81 \\
Subdataset 14 & 2021-06-04 & 2023-06-05 & 486 & 2023-06-06 & 2023-09-28 & 81 & 2023-10-09 & 2024-01-30 & 81 \\
\hline
\end{tabular}
}
\end{table}

Moreover, each subdataset underwent a filtering process for the stock pool selection:
\begin{enumerate}
    \item Stocks that were delisted, had negative net assets, were newly issued, or were under risk warning were excluded.
    \item Stocks that were suspended from trading or hit the price limit on the trading day were also excluded.
\end{enumerate}

\subsection{Factor Construction}
This study constructs price-volume factors for each subdataset by mining daily market data. Utilizing the Alpha360 factor library developed by Microsoft's Qlib framework, we constructed 360 price-volume factors, categorized into six types, with each type comprising 60 factors. Table~\ref{tab:factor_construction} presents examples of factor construction methods for the first two factors in each category.

\begin{table}[h]
\centering
\caption{Construction Methods for Six Categories of Price-Volume Factors}
\label{tab:factor_construction}
\small 
\begin{tabular}{ccc}
\toprule
\textbf{Factor Category} & \textbf{Factor Name} & \textbf{Price-Volume Factor Construction Method} \\
\midrule
Close/Close & CLOSE1 & $\text{Ref}(\text{close},1)/\text{close}$ \\
 & CLOSE2 & $\text{Ref}(\text{close},2)/\text{close}$ \\
 & \dots & \dots \\
\midrule
Open/Close & OPEN1 & $\text{Ref}(\text{open},1)/\text{close}$ \\
 & OPEN2 & $\text{Ref}(\text{open},2)/\text{close}$ \\
 & \dots & \dots \\
\midrule
High/Close & HIGH1 & $\text{Ref}(\text{high},1)/\text{close}$ \\
 & HIGH2 & $\text{Ref}(\text{high},2)/\text{close}$ \\
 & \dots & \dots \\
\midrule
Low/Close & LOW1 & $\text{Ref}(\text{low},1)/\text{close}$ \\
 & LOW2 & $\text{Ref}(\text{low},2)/\text{close}$ \\
 & \dots & \dots \\
\midrule
Vwap/Close & VWAP1 & $\text{Ref}(\text{vwap},1)/\text{close}$ \\
 & VWAP2 & $\text{Ref}(\text{vwap},2)/\text{close}$ \\
 & \dots & \dots \\
\midrule
Volume/Volume & VOLUME1 & $\text{Ref}(\text{volume},1)/(\text{volume}+1e-12)$ \\
 & VOLUME2 & $\text{Ref}(\text{volume},2)/(\text{volume}+1e-12)$ \\
 & \dots & \dots \\
\bottomrule
\end{tabular}
\end{table}

The function \texttt{Ref} retrieves the historical value of a variable. For instance, in the Close/Close category, the price-volume factors CLOSE0-CLOSE59 are calculated as the ratio of the close value from 0 to 59 periods back to the current close value. The construction of other categories follows a similar rationale. This study constructs these 360 price-volume factors daily for every stock in the stock pool.

\subsection{Factor Neutralization and Factor Value Preprocessing}

\subsubsection{Factor Value Preprocessing}
For each subdataset divided, factor values corresponding to each trading day are preprocessed, which includes the following steps:
\begin{enumerate}
    \item \textbf{Missing Values:} Fill missing factor values using the information from the previous trading day of the stock.
    \item \textbf{Extreme Value Treatment:} Exclude data points that fall beyond three standard deviations from the mean, assuming a normal distribution.
    \item \textbf{Standardization:} Apply Z-score normalization to each factor sequence to standardize the values.
\end{enumerate}

\subsubsection{Neutralization Process}
Industry and market value neutralization are employed to eliminate the influence of a stock's industry and market value. The process for each subdataset is as follows:
\begin{enumerate}
    \item Convert each stock's industry category into dummy variables.
    \item Conduct a linear regression where the dependent variable is the factor sequence, and the independent variables include the industry dummy variables and the market value variable.
    \item Use the residual series from the regression as the neutralized factor sequence.
\end{enumerate}

\subsection{Factor Effectiveness Testing}

The Information Coefficient (IC) is a statistical metric used to assess the correlation between a factor and asset returns. In this study, the Spearman rank correlation coefficient is utilized to calculate the IC value, referred to as RankIC.

For instance, consider the subdataset from the period of May 9, 2023, to August 2, 2023, for IC value analysis. For each trading day within this period, IC values can be calculated for the 360 price-volume factors. The following discussion presents an example of IC values for 12 specific price-volume factors as shown in Table \ref{tab:factor_ic_analysis}.

\begin{table}[h]
\centering
\caption{Factor IC Value Analysis}
\label{tab:factor_ic_analysis}
\resizebox{\textwidth}{!}{
\begin{tabular}{lcccc}
\toprule
\textbf{Factor Name} & \textbf{Mean IC} & \textbf{IC Standard Deviation} & \textbf{Proportion of IC \textgreater 0 (\%)} & \textbf{Proportion of $|\mathrm{IC}|$ \textgreater 0.02 (\%)} \\
\midrule
CLOSE1 & 0.098 & 0.038 & 99.61 & 98.82 \\
CLOSE30 & 0.114 & 0.080 & 89.41 & 91.76 \\
OPEN1 & 0.135 & 0.053 & 98.82 & 97.25 \\
OPEN30 & 0.113 & 0.079 & 90.20 & 92.16 \\
HIGH1 & 0.153 & 0.079 & 97.25 & 96.47 \\
HIGH30 & 0.118 & 0.076 & 92.55 & 92.55 \\
LOW1 & 0.089 & 0.075 & 87.84 & 89.80 \\
LOW30 & 0.120 & 0.085 & 87.45 & 89.41 \\
VWAP1 & 0.121 & 0.042 & 99.61 & 99.22 \\
VWAP30 & 0.080 & 0.902 & 90.20 & 91.76 \\
VOLUME1 & 0.102 & 0.060 & 93.33 & 90.98 \\
VOLUME30 & 0.079 & 0.093 & 78.82 & 92.94 \\
\bottomrule
\end{tabular}
}
\end{table}

In multifactor stock selection models, an Information Coefficient (IC) value greater than 0.02 is generally considered indicative that the factor partially reflects future stock trends, hence deemed effective. As demonstrated in Table \ref{tab:factor_ic_analysis}, the listed twelve factors all have mean IC values greater than 0.07, with the proportion of IC absolute values exceeding 0.02 being over 89\%, affirming their effectiveness. Additionally, these factors exhibit over 85\% positivity, indicating their predictions align with market trends most of the time.

For other price-volume factors in this subdataset, and factors in other subdatasets, IC values are analyzed in the same manner. If a factor's mean IC is below 0.02, it is considered ineffective and excluded from further analysis.


\section{Experiments}
This paper conducts a series of experiments to explore the performance of the Stockformer model in predicting stock market returns and trend directions. Detailed analyses are carried out addressing the following five research questions:
\subsection{Research Questions}

\begin{enumerate}
    \item Does the Stockformer model outperform existing baseline models in predicting returns and market trends?
    \item What impact do different components of the Stockformer model (such as the time attention mechanism, graph embedding techniques, and multi-supervision learning) have on the model's performance?
    \item How do hyperparameters affect the performance of Stockformer?
    \item During the backtesting phase, considering the probability predictions of trend directions and return forecasts may lead to different backtesting performances, which of these two outputs should be chosen to maximize potential investment returns when labels are unknown (unable to determine evaluation metrics)?
    \item In the investment strategy backtesting phase, does the Stockformer model outperform existing baseline models?
\end{enumerate}

\subsection{Experimental Setup}
\subsubsection{Dataset}
As previously described in subsection \ref{subsec:data_sources_stock_selection}, this section utilizes stocks from the CSI 300 Index, which includes the top 300 stocks in terms of market liquidity and capitalization, reflecting the overall trend of the A-share market. For the dataset, we use a matrix format where the rows represent dates and the columns represent the return series of each of the 300 stocks. This serves as the input for wavelet transformations. Based on the sign of the return rate at opening, binary up/down trend indicators (0 or 1) are generated. Alongside, 360 price-volume factors are constructed according to the methods outlined in section \ref{sec:factor_construction}, forming the multifactor input for our model. Thus, the input dimensionality for our model is $\mathbb{R}^{T_1 \times N \times 362}$.

The model uses data from the first 20 time steps ($T_1 = 20$) to predict the returns and stock price trends for the next two time steps ($T_2 = 2$). These datasets are divided into training (75\%), validation (12.5\%), and testing sets (12.5\%), in chronological order and according to the aforementioned day counts.

\subsubsection{Evaluation Metrics}
The evaluation metrics used in this study are divided into two main parts to comprehensively assess the performance of the Stockformer model in stock market forecasting. The first part evaluates the model's predictive performance, focusing on its accuracy and efficiency in forecasting stock returns and price trends. The second part evaluates the investment portfolio during the backtesting phase of the investment strategy, involving the use of historical data to test the effectiveness of model predictions in actual investment operations, assessing its performance and risk management capabilities. This comprehensive evaluation aims to showcase the superiority of the Stockformer model in stock return prediction and investment strategy formulation.

\paragraph{\textbf{Predictive Performance Evaluation Metrics}}
To comprehensively assess model performance across different prediction tasks, this study divides the evaluation metrics into two categories: return prediction metrics and trend prediction metrics.

\subparagraph{Return Prediction Metrics}
The following metrics are used to evaluate model performance in the return prediction phase:
\begin{enumerate}
    \item \textbf{IC (Information Coefficient)}\citep{lin2021learning}: The IC is a statistical measure that assesses the correlation between a single predictor and actual asset returns. It is calculated using the Spearman correlation coefficient, reflecting the correlation between the rankings of factors and returns. The formula is as follows:
    \begin{equation} \label{eq:ic}
    \mathrm{IC}=\frac{1}{N} \frac{(\hat{\mathbf{y}}-\operatorname{mean}(\hat{\mathbf{y}}))^{\mathrm{T}}(\mathbf{y}-\operatorname{mean}(\mathbf{y}))}{\operatorname{std}(\hat{\mathbf{y}}) \cdot \operatorname{std}(\mathbf{y})}
    \end{equation}
    \item \textbf{ICIR (Information Coefficient Information Ratio)}\citep{lin2021learning}: ICIR measures the stability and consistency of IC values, akin to the Sharpe ratio of IC, and is used to evaluate the reliability of factor performance. The formula is as follows:
    \begin{equation} \label{eq:icir}
    \mathrm{ICIR} = \frac{\operatorname{mean}(\mathrm{IC})}{\operatorname{std}(\mathrm{IC})}
    \end{equation}
    \item \textbf{RankIC}\citep{li2019individualized}: RankIC is an information coefficient based on ranking, using the Spearman rank correlation coefficient instead of raw values, which is more resistant to outliers. The formula is as follows:
    \begin{equation} \label{eq:rankic}
    \mathrm{RankIC}=\frac{1}{N} \frac{(\operatorname{rank}(\hat{\mathbf{y}})-\operatorname{mean}(\operatorname{rank}(\hat{\mathbf{y}})))^{\mathrm{T}}(\operatorname{rank}(\mathbf{y})-\operatorname{mean}(\operatorname{rank}(\mathbf{y})))}{\operatorname{std}(\operatorname{rank}(\hat{\mathbf{y}})) \cdot \operatorname{std}(\operatorname{rank}(\mathbf{y}))}
    \end{equation}
    \item \textbf{RankICIR}\citep{li2019individualized}: Information ratio of RankIC, calculated similarly to ICIR but applied to RankIC values, used to assess the stability and predictive power of the rank-based information coefficient. The formula is as follows:
    \begin{equation} \label{eq:rankicir}
    \mathrm{RankICIR} = \frac{\operatorname{mean}(\mathrm{RankIC})}{\operatorname{std}(\mathrm{RankIC})}
    \end{equation}
\end{enumerate}

\subparagraph{Trend Prediction Metrics}
Used to evaluate model accuracy in predicting stock price trends:
\begin{enumerate}
    \item \textbf{Directional Accuracy}: This metric assesses the model's accuracy in predicting the direction of stock price movements. For the Stockformer model, directional accuracy is calculated using the trend classification output; for baseline models, it is determined by the sign (positive or negative) of the predicted returns. The formula is as follows:
    \begin{equation} \label{eq:directional_accuracy}
    \text{Directional Accuracy} = \frac{\text{Number of Correct Predictions}}{\text{Total Number of Predictions}} \times 100\%
    \end{equation}
\end{enumerate}

\paragraph{\textbf{Portfolio Performance Metrics}}
This section introduces four key metrics used to assess the performance of investment portfolios, aiding investors in better understanding the risk and return characteristics of their investments.

\begin{enumerate}
    \item \textbf{Annualized Return}: The annualized return helps investors comprehend the long-term performance of an investment portfolio, reflecting the time value of the investment. The formula is as follows:
    \begin{equation} \label{eq:annualized_return}
    R_y = (1 + R)^{250 / (T_2 - T_1)}
    \end{equation}
    where $R$ is the total return, $T_1$ is the initial investment date, and $T_2$ is the closing date.

    \item \textbf{Maximum Drawdown}: Maximum drawdown measures the largest loss in a portfolio during a specific period and is used to assess the level of risk under market fluctuations. The formula is as follows:
    \begin{equation} \label{eq:max_drawdown}
    \text{Max Drawdown} = \frac{P - Q}{P}
    \end{equation}
    where $P$ is the highest net value during the period, and $Q$ is the lowest net value following the peak.

    \item \textbf{Annualized Volatility}: Annualized volatility measures the variability of asset prices or an investment portfolio over a given time period. The formula is as follows:
    \begin{equation} \label{eq:annualized_volatility}
    \text{Volatility} = \sqrt{252} \times \sigma_R
    \end{equation}
    where $\sigma_R$ is the standard deviation of daily returns.

    \item \textbf{Sharpe Ratio }\citep{sharpe1966mutual}: The Sharpe Ratio evaluates the return of an investment portfolio relative to the risk it has taken on. The formula is as follows:
    \begin{equation} \label{eq:sharpe_ratio}
    S_p = \frac{r_p - r_f}{\sigma_P}
    \end{equation}
    where $r_p$ is the return of the portfolio, $r_f$ is the risk-free rate (such as the yield on one-year government bonds), and $\sigma_P$ is the standard deviation of the portfolio's returns.
\end{enumerate}

\subsubsection{Baselines}
In this study, our proposed model, Stockformer, is compared with the following benchmark models, which were introduced in the related work section. In total, ten baseline models are utilized for comparison:
\begin{enumerate}
    \item XGBoost\citep{chen2016xgboost}: An optimized decision tree model using gradient boosting techniques, widely used in classification and regression tasks.
    \item LightGBM\citep{ke2017lightgbm}: A framework based on gradient boosting that is characterized by fast speed, high efficiency, and friendliness to large-scale data.
    \item CatBoost\citep{dorogush2018catboost}: A high-performance gradient boosting tree model that automatically handles categorical features, reducing the need for pre-processing before model training.
    \item LSTM\citep{hochreiter1997long}: A recurrent neural network capable of learning long-term dependencies, widely applied in time series forecasting and natural language processing.
    \item GRU\citep{cho2014learning}: A simplified version of LSTM, offering similar performance but with fewer parameters and faster training.
    \item ALSTM\citep{wang2020alstm}: Enhances LSTM by incorporating an attention mechanism, which improves the recognition of key information.
    \item TCN\citep{bai2018empirical}: A network based on causal convolutions, effective in handling time series data, suitable for scenarios requiring extensive historical information.
    \item GATs\citep{velickovic2017graph}: Graph Attention Networks that allow nodes to dynamically compute their importance based on the features of their neighbors through a graph attention mechanism.
    \item Transformer\citep{vaswani2017attention}: Based on self-attention mechanisms, capable of capturing global dependencies, effectively processing sequence data.
    \item Localformer\citep{zheng2023preserving}: By utilizing a Hilbert curve to unfold the image matrix, it better preserves the smoothness of local information, suitable for images and other two-dimensional data.
\end{enumerate}

\subsubsection{Parameter Settings}

The Stockformer model was implemented using the PyTorch framework and trained with the Adam optimizer for a total of 100 epochs. Due to computational resource constraints, the batch size for input data was set to 2. Within the Stockformer model, the number of heads $e$ in the attention mechanism was set to 1, and the base dimension $d_e$ was set to 128. The number of layers $L$ in the spatiotemporal encoder was set to 2. To capture high-frequency temporal dependencies, dilated causal convolution layers with a kernel size of $J=2$ were stacked. The initial learning rate was set at 0.001, with a decay rate of 0.1 for adjusting the learning rate. Dropout was also employed, with a dropout rate of 0.2, to mitigate the risk of overfitting.

\subsubsection{Computational Environment}

The computational setup for this study is detailed below:
\begin{enumerate}
    \item \textbf{CPU:} Utilized two Intel(R) Xeon(R) Platinum 8352V CPUs @ 2.10GHz, each with 32 cores and 64 threads, totaling 64 cores and 128 threads, providing substantial parallel computing power. CPU clock speeds ranged from 800 MHz to 3500 MHz, with support for Intel VT-x virtualization technology.
    \item \textbf{GPU:} The system was equipped with an NVIDIA GeForce RTX 4090, boasting 24564 MB of video memory.
    \item \textbf{Memory:} The system included high-speed cache, comprising 3 MiB of L1d cache, 2 MiB of L1i cache, 80 MiB of L2 cache, and 108 MiB of L3 cache, ensuring efficient data processing.
    \item \textbf{Operating System and Framework:} All model training was conducted on a Linux system based on the X86\_64 architecture, utilizing the PyTorch deep learning framework.
\end{enumerate}

This configuration provides an efficient and stable training environment for deep learning models.

\subsection{Comparative Analysis of Predictive Performance (RQ 1)}
This section will detail the prediction results of the Stockformer model on the test sets of 14 sub-datasets for stock return sequences and trend movements. We will compare the Stockformer model with several advanced stock return prediction models identified in the research. These baseline models are categorized into the following groups: the first category includes ensemble learning models such as XGBoost, LightGBM, and CatBoost; the second category comprises recurrent neural network models including LSTM, GRU, and ALSTM; the third category consists of component models of our system, namely Temporal Convolutional Networks (TCN) and Graph Attention Networks (GATs); the fourth category encompasses models with an Encoder-Decoder architecture that incorporate self-attention mechanisms, such as Transformer and Localformer. Through this categorized comparison, we aim to thoroughly explore the performance advantages of the Stockformer model across various evaluation metrics. The research findings are presented in Table \ref{tab:model_performance}, which includes average values of various prediction performance metrics used across the 14 sub-datasets, thereby visually demonstrating the comparative performance of Stockformer against other state-of-the-art (SOTA) models.

\begin{table}[h]
\centering
\caption{Model Predictive Performance Comparison: \textbf{Bold} indicates the best performance, \underline{underlined} indicates the second best. An upward arrow (\(\uparrow\)) signifies that higher metric values indicate better model performance, while a downward arrow (\(\downarrow\)) indicates that lower metric values are better. The table lists the average values of various prediction performance metrics for the test set (out-of-sample predictions) on the 14 subdatasets.}
\label{tab:model_performance}
\begin{tabular}{lcccccc}
\toprule
Model        & IC\({\uparrow}\) & Rank IC\({\uparrow}\) & ICIR\({\uparrow}\) & Rank ICIR\({\uparrow}\) & Directional Accuracy\({\uparrow}\)(\%) \\
\midrule
XGBoost      & -0.0077 & -0.0104 & -0.0594 & -0.0788   & 51.11                     \\
LightGBM     & -0.0070 & -0.0202 & -0.0531 & -0.1572   & 54.11                     \\
CatBoost     & -0.0214 & -0.0269 & -0.1609 & -0.2135   & 52.25                     \\
LSTM         & -0.0014 & -0.0038 & -0.0168 & -0.0372   & 53.04                     \\
GRU          & 0.0068  & 0.0076  & 0.0637  & \underline{0.0674}    & 52.17                     \\
ALSTM        & \underline{0.0124}  & \underline{0.0081}  & \underline{0.0952}  & 0.0603    & 52.85                     \\
TCN          & 0.0019  & -0.0004 & 0.0190  & 0.0015    & 53.83                     \\
GATs         & 0.0108  & 0.0059  & 0.0839  & 0.0475    & 54.52                     \\
Transformer  & 0.0061  & 0.0048  & 0.0511  & 0.0418    & 49.29                     \\
Localformer  & 0.0071  & 0.0104  & 0.0646  & 0.0928    & \underline{54.87}         \\
\midrule
\textbf{Stockformer}  & \textbf{0.0294} & \textbf{0.0344} & \textbf{0.1921} & \textbf{0.2669} & \textbf{57.46} \\
\bottomrule
\end{tabular}
\end{table}

The table \ref{tab:model_performance} clearly demonstrates that Stockformer excels across multiple key performance indicators. Notably, it significantly outperforms other models on the IC and ICIR metrics, indicating a strong positive correlation and high stability and consistency between its predictive factors and actual asset returns. Additionally, it achieves the highest Directional Accuracy at 57.46\%, substantially surpassing other models, further validating its accuracy and reliability in predicting market trends.

When comparing the groups of baseline models, we observe the following characteristics:
\begin{enumerate}
    \item The ensemble learning models (XGBoost, LightGBM, and CatBoost) have a speed advantage when processing large datasets, but generally fall short in capturing complex market dynamics compared to deep learning models. Although these models perform reasonably well in Directional Accuracy, their poor performance on IC and ICIR metrics indicates limitations in prediction accuracy and stability.
    \item Recurrent neural network models (LSTM, GRU, and ALSTM) exhibit strong performance in capturing the temporal dependencies in time series data. Among these, ALSTM stands out in all baseline models, particularly on the IC and ICIR metrics, demonstrating high predictive consistency and stability. The inclusion of an attention mechanism substantially contributes to ALSTM’s impressive performance, enhancing its ability to effectively process and parse critical information in time series data, especially in complex and long-term dependencies. The addition of the attention layer not only helps the model focus on the historical information most critical to future predictions but also improves the model’s adaptability to dynamic changes in the data. Furthermore, ALSTM’s performance reaffirms the correctness of our multiple uses of attention mechanisms in the design of Stockformer.
    \item Models utilizing Encoder-Decoder architectures with integrated self-attention mechanisms, such as Transformer and Localformer, are theoretically expected to excel in capturing long-distance dependencies. In terms of performance metrics for rate of return predictions, these models demonstrate some stability, yet they do not exhibit superior advantages. Notably, in the performance metrics for trend prediction accuracy, Transformer performs poorly, revealing potential limitations in its ability to predict actual financial market trends. This suggests that when employing these models, a greater emphasis may be required on model tuning and feature adaptation to specific scenarios.
    \end{enumerate}

\paragraph{\textbf{Advantages of Stockformer}}
The exceptional performance of Stockformer can be attributed to its unique architecture that integrates various powerful deep learning technologies, thereby optimizing the model's ability to process complex stock market data. Specifically:

\begin{enumerate}
  \item \textbf{Temporal Convolutional Networks (TCN)} provide effective capture of long-term dependencies in time series data.
  \item \textbf{Graph Attention Mechanisms (GATs)} enable the model to consider interactions and mutual influences among stocks, enhancing its understanding of market dynamics.
  \item \textbf{Self-Attention Mechanisms (Transformer and Localformer)} strengthen the model's ability to capture complex interactions among various points in time series data.
  \item \textbf{Multi-Supervision Learning}: Through a sophisticated multi-task learning framework, Stockformer can simultaneously perform classification and regression tasks, allowing classification and regression to supervise each other and perform multi-dimensional predictions on stock trends and returns, enhancing the model's adaptability and predictive accuracy.
\end{enumerate}

In subsequent model backtesting and comparisons, this study selected the best-performing models from each baseline category --- LightGBM, ALSTM, TCN, GATs, Localformer, and our proprietary Stockformer model --- for detailed analysis.

\subsection{Ablation Study (RQ 2)}

To investigate the effects of different components within the Stockformer model, this study compares it against six distinct variants:
\begin{enumerate}
\item "w/o D (Decoupling Flow Layer)": Stockformer without the Decoupling Flow Layer;
\item "w/o T (Dilated causal convolution and Temporal Attention)": Stockformer lacking both Dilated Causal Convolution and Temporal Attention;
\item "w/o G (Graph)": Stockformer devoid of spatial and temporal graph embeddings;
\item "w/o F (Fusion)": Stockformer where Fusion Attention is replaced by addition operations;
\item "w/o Reg": Stockformer that outputs only the classification task, using only the classification results as the supervisory signal;
\item "w/o Cla": Stockformer that outputs only the regression task, using only the regression results as the supervisory signal.
\end{enumerate}

\begin{table}[htbp]
\centering
\caption{Stockformer Ablation Study: \textbf{Bold} indicates the best performance, \underline{underlined} indicates the second-best performance. An upward arrow (\(\uparrow\)) indicates that higher metric values denote better model performance, while a downward arrow (\(\downarrow\)) indicates that lower metric values are preferable. The table lists the average values of various prediction performance metrics for the test set (out-of-sample predictions) on the 14 subdatasets.}
\label{tab:stockformer_variants_comparison}
\begin{tabular}{lccccc}
\toprule
Variant & IC(\(\uparrow\)) & Rank IC(\(\uparrow\)) & ICIR(\(\uparrow\)) & Rank ICIR(\(\uparrow\)) & Directional Accuracy(\(\uparrow\))(\%) \\
\midrule
w/o D & 0.0122 & 0.0202 & 0.0504 & 0.0663 & 52.81 \\
w/o T & 0.0122 & 0.0205 & 0.0506 & 0.0840 & 52.65 \\
w/o G & 0.0104 & 0.0197 & 0.0498 & 0.0742 & 51.83 \\
w/o F & 0.0186 & 0.0267 & 0.0641 & 0.0928 & 50.30 \\
w/o Reg & 0.0205 & 0.0281 & 0.0888 & 0.1005 & \underline{54.60} \\
w/o Cla & \underline{0.0237} & \underline{0.0291} & \underline{0.1233} & \underline{0.1215} & 53.32 \\
\midrule
\textbf{Stockformer}  & \textbf{0.0294} & \textbf{0.0344} & \textbf{0.1921} & \textbf{0.2669} & \textbf{57.46} \\
\bottomrule
\end{tabular}
\end{table}

The data in Table~\ref{tab:stockformer_variants_comparison} clearly demonstrate that the complete Stockformer model outperforms its various variants, highlighting the importance of each component within the model. Removing or modifying key components such as the Decoupling Flow Layer, Dilated Causal Convolution, Graph Embeddings, or Fusion Attention leads to a noticeable decline in performance. Particularly, the removal of Graph Embeddings significantly impacts the model's accuracy and error metrics, underscoring the critical role of modeling complex data relationships for prediction accuracy. Furthermore, Stockformer's strategy of Multi-Supervision, which involves handling both classification and regression tasks concurrently, enhances the model's generalization capabilities and adaptability. This multi-task learning approach not only provides the model with a richer learning signal but also aids in capturing and understanding the complexities of data across multiple dimensions, thereby improving overall predictive performance.

\subsection{Hyperparameter Sensitivity Analysis (RQ 3)}

\begin{figure}[H]
    \centering
    \includegraphics[width=1\textwidth]{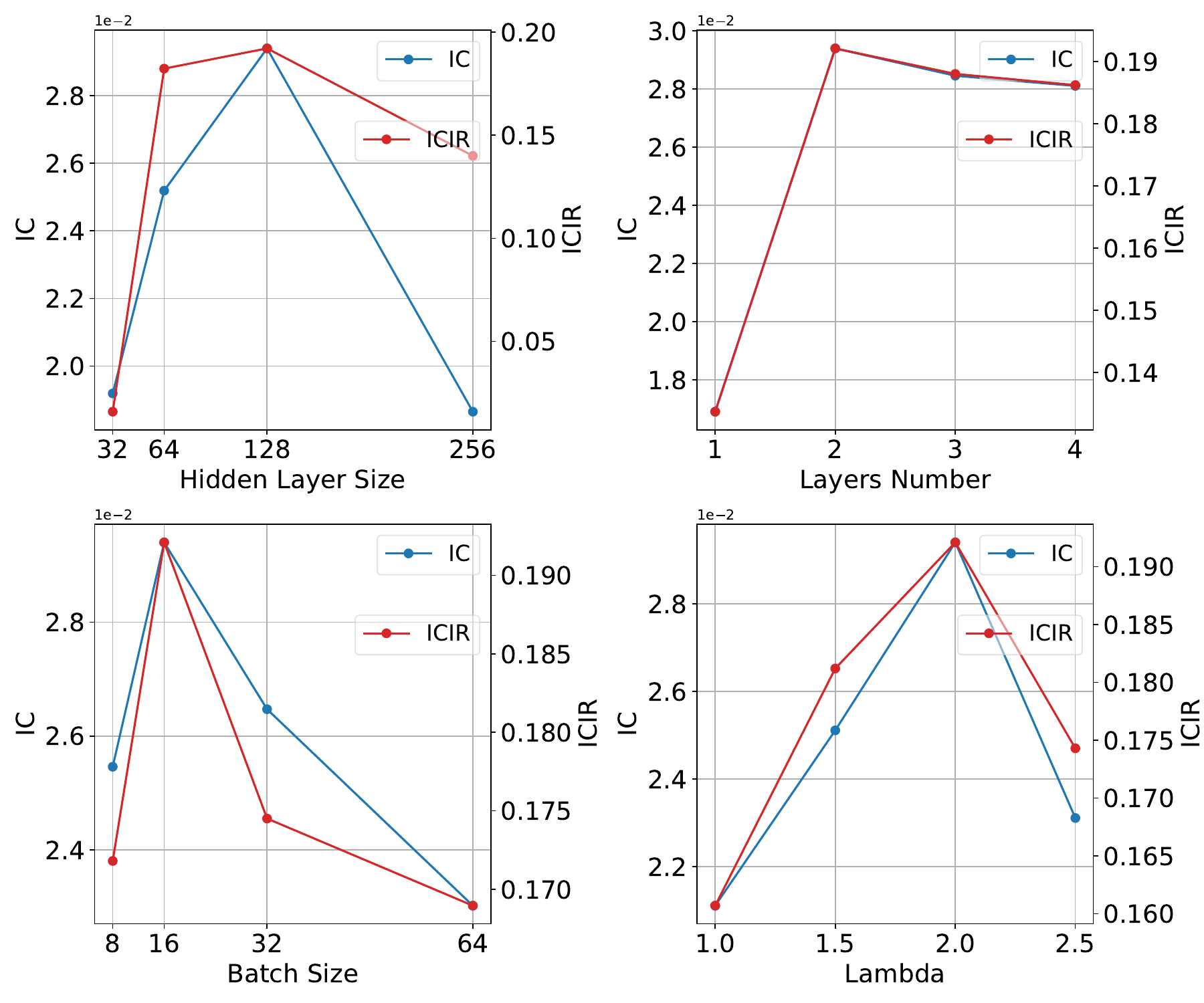}
    \caption{Hyperparameter Sensitivity Analysis. In the figure, the blue line represents the Information Coefficient (IC) values on the test set (out-of-sample predictions), while the red line represents the Information Coefficient Information Ratio (ICIR) values. The metrics evaluate the impact of various hyperparameters including hidden layer size, number of encoder layers, batch size, and classification loss weight on the model's prediction accuracy and stability.}
    \label{fig:Hyper} 
\end{figure}

In this study, we conducted a sensitivity analysis of hyperparameters for the Stockformer model, covering hidden layer sizes (32, 64, 128, 256), number of encoder layers (1 to 4), batch sizes (8, 16, 32, 64), and classification loss weights (1.0, 1.5, 2.0, 2.5), as shown in Figure~\ref{fig:Hyper}. The various settings of these hyperparameters were thoroughly examined for their impact on key performance indicators such as Information Coefficient (IC) and Information Coefficient Information Ratio (ICIR), with the aim of identifying optimal configurations to enhance the model's prediction accuracy and stability.

The analysis results indicate that optimal performance of the Stockformer is achieved when the hidden layer size is set to 128 and the batch size to 16. These findings highlight the ideal combination of model parameters under specific settings, and how further increasing or decreasing these values may lead to performance degradation. The best performance is observed when the number of encoder layers is set to 2, suggesting that a moderate increase in model depth can effectively enhance performance, but additional layers may not yield further benefits. In adjusting the weight for the classification task, a value of 2.0 shows the best performance, emphasizing the importance of appropriately adjusting loss weights in balancing classification and regression tasks.

\subsection{Backtesting Input Selection (Research Question 4)}
\label{subsec:backtest_input_selection}

In the backtesting phase of stock market prediction models (forecasting future outcomes), selecting the appropriate output of Stockformer—either classification or regression results—for backtesting is crucial. This study compares the backtesting performance of different outputs—the probability of trend predictions (classification output) and the results of return predictions (regression output)—to determine which output can maximize potential investment returns. The following analysis and recommendations are provided:

\paragraph{\textbf{High Confidence Prediction Proportion Analysis}}
The proportion of high confidence predictions is a metric used to measure the certainty of model predictions, defined in this study as the proportion of upward trend predictions (label 1) with a probability value greater than 0.6 or less than 0.4. This proportion reflects the level of certainty in the model's predictions. A higher proportion indicates a higher certainty in the model's classification predictions, while a lower proportion indicates greater uncertainty. Based on the backtesting results from 14 sub-datasets created by sliding window segmentation (detailed backtesting results are plotted in the next subsection), we obtain Table~\ref{tab:confidence_results}, comparing the performance of regression and classification outputs across each sub-dataset.

\begin{table}[htbp]
\centering
\caption{High Confidence Prediction Proportion and Best Outcome Analysis}
\label{tab:confidence_results}
{\footnotesize 
\begin{tabular}{lcc}
\toprule
Backtesting Date Range & High Confidence Prediction Proportion & Best Result \\
\midrule
2018-03-01 -- 2020-10-29 & 0.6886 & cla \\
2018-05-31 -- 2021-01-25 & 0.6797 & cla \\
2018-08-27 -- 2021-04-28 & 0.2558 & cla \\
2018-11-28 -- 2021-07-28 & 0.1139 & reg \\
2019-03-04 -- 2021-11-01 & 0.0388 & reg \\
2019-06-03 -- 2022-01-26 & 0.2422 & cla \\
2019-08-28 -- 2022-05-05 & 0.6407 & cla \\
2019-11-29 -- 2022-08-01 & 0.3771 & cla \\
2020-03-04 -- 2022-11-02 & 0.3183 & cla \\
2020-06-03 -- 2023-02-03 & 0.7132 & cla \\
2020-08-31 -- 2023-05-05 & 0.0292 & reg \\
2020-12-02 -- 2023-08-02 & 0.2599 & cla \\
2021-03-05 -- 2023-11-03 & 0.1180 & reg \\
2021-06-04 -- 2024-01-30 & 0.1297 & reg \\
\bottomrule
\end{tabular}
} 
\end{table}

\paragraph{\textbf{Model Selection Recommendations}}
Based on the analysis of high confidence prediction proportions and optimal outcomes over different time ranges, we offer the following recommendations:

\begin{enumerate}
    \item \textbf{When high confidence prediction proportion is high}: Classification outputs generally provide better results. For instance, when the proportion of high confidence predictions exceeds 20\%, classification outcomes often demonstrate higher prediction accuracy. This may be attributed to the classification model's effectiveness in capturing market trends under these conditions.
    \item \textbf{When high confidence prediction proportion is low}: Regression outputs may be a more suitable choice. In scenarios where prediction confidence is low, regression models can better handle the uncertainty and volatility in the data, providing more continuous and stable outputs.
\end{enumerate}

These strategies can assist investors and analysts in maximizing investment returns during the backtesting phase when using Stockformer's results by selecting the most appropriate model output.


\subsection{Strategy Backtesting}

\begin{figure}[!htbp]
	\centering
	\includegraphics[width=0.8\columnwidth]{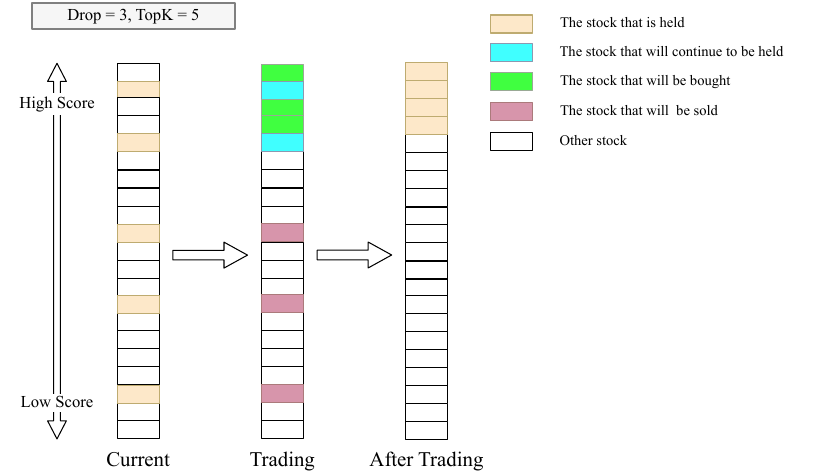} 
	\caption{In the TopKDropout strategy, with TopK as 5 and Drop as 3: Prior to the subsequent position adjustment, the three lowest-ranked stocks are discarded. From the remaining stocks, excluding the current holdings, the top three with the highest scores (given by TopK-Drop, i.e., 5-3) are chosen for the next trading cycle.} 
	\label{fig:topkdropout} 
\end{figure}

\subsubsection{Topk-Dropout Strategy}
The Topk-Dropout strategy is a ranking-based quantitative stock selection strategy that constructs a portfolio by ranking stocks based on their factor values and retaining only the top k stocks to achieve stock selection. Specifically, the Topk-Dropout strategy includes the following steps:
\begin{enumerate}
    \item On the current trading day, calculate the factor values for each stock given a set of factors.
    \item Sort the stocks by factor values, sell stocks that are ranked beyond the top k in prediction scores, and buy an equal number of stocks that are newly ranked in the top k, ensuring that exactly k stocks are held at all times, except at the start of trading when this number is zero.
    \item Allocate funds to the selected stocks equally by weight and conduct buying and selling according to the trading frequency.
    \item At the next rebalancing period, re-sort the stocks by factor values and select stocks, repeating the steps above.
\end{enumerate}

The core idea of the Topk-Dropout strategy is to select stocks that rank high in factor values, assuming these stocks are more likely to perform well in the future. Unlike other stock-picking strategies, the TopK-Dropout strategy focuses on investing in only the top k ranked stocks and discards the rest. This method enhances the concentration of the investment portfolio, reduces transaction costs, and also decreases the number of low-quality stocks in the portfolio, thus improving the effectiveness and stability of the stock-picking strategy.

Figure \ref{fig:topkdropout} shows an example of the TopK-Dropout strategy where k is set to 5. Before the next position adjustment, the three lowest-ranked stocks are discarded. From the remaining stocks, excluding the current holdings, the top k stocks with the highest scores are chosen for the next trading cycle. The investment strategy settings adopted in this paper are as follows:
\begin{enumerate}
    \item The period from 2018 to 2024 is divided into 14 time intervals, with independent backtesting conducted on each interval.
    \item The trading strategy chosen is TopK-Dropout, with k set to 5.
    \item The stock pool is selected from the screened constituents of the CSI 300 index.
    \item The benchmark for comparison is the CSI 300 index.
    \item The trading frequency is daily, with the model predicting the T+1 day return rate, i.e., the price change from the T+1 day closing price to the T+2 day closing price.
    \item The transaction cost is 0.1\% per side, with stamp duty at 0.1\% before August 27, 2023, and 0.05\% thereafter.
\end{enumerate}

\subsubsection{Investment Strategy Backtesting}

\begin{figure}[!htbp]
	\centering
	\includegraphics[width=\columnwidth]{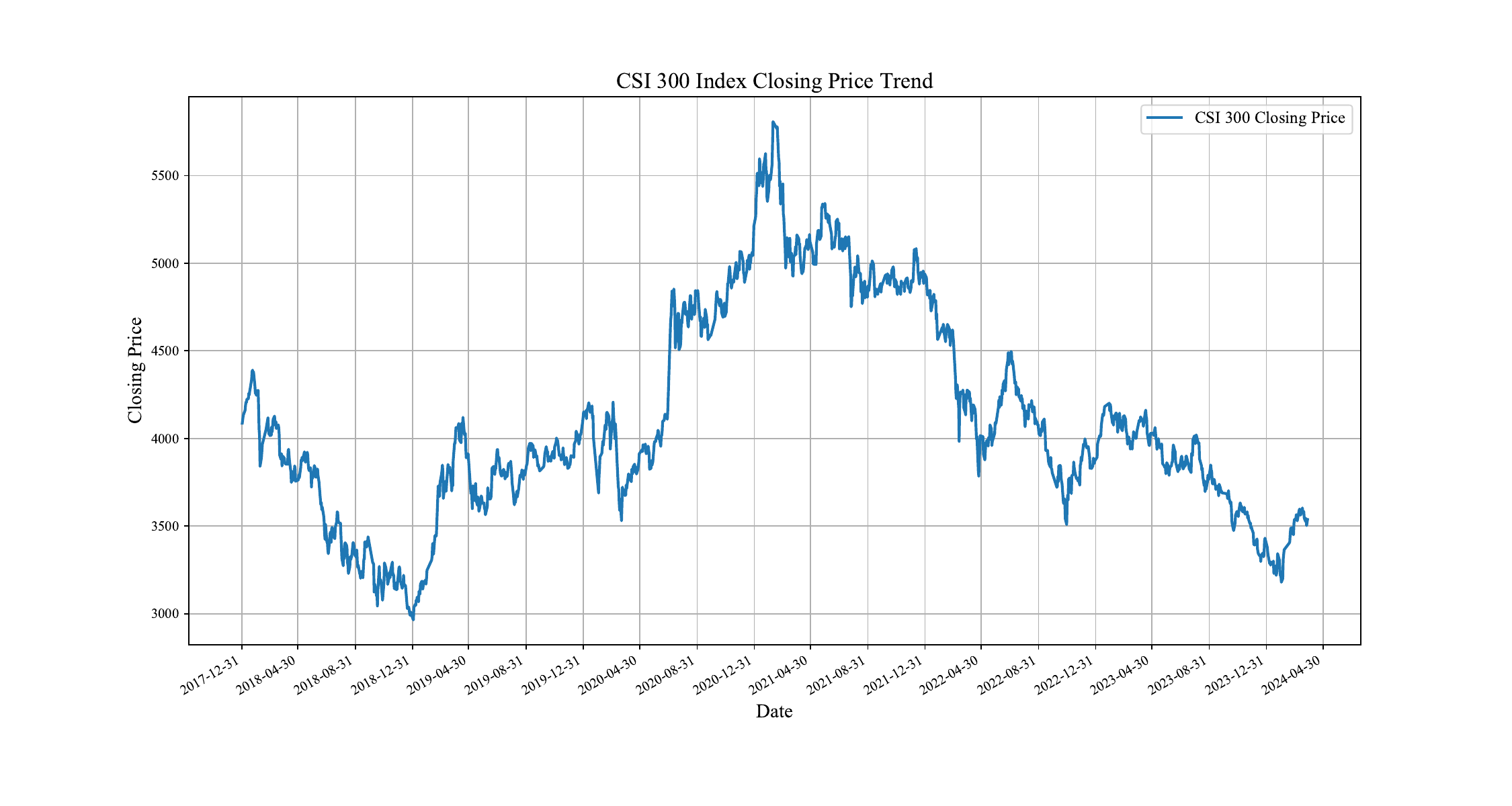} 
	\caption{The Trend of the CSI 300 Index from 2018 to 2024} 
	\label{fig:HS300_Close_Price_Trend} 
\end{figure}

Figure \ref{fig:HS300_Close_Price_Trend} shows the trend of the CSI 300 Index from 2018 to 2024, which largely reflects the overall market trend. During this period, the market exhibited phases of rise, fall, and fluctuation. The stability of investment strategies under different market conditions is a critical issue that needs attention.

\begin{table}[htbp]
\centering
\caption{Selection of Backtesting Intervals}
\label{tab:backtesting_intervals}
\begin{tabular}{lcc}
\toprule
Backtesting Start Date & Backtesting End Date & Market Condition \\
\midrule
2020-11-02 & 2021-01-25 & Uptrend \\
2022-01-28 & 2022-05-05 & Downtrend \\
2023-05-09 & 2023-08-02 & Fluctuation \\
\bottomrule
\end{tabular}
\end{table}

This study selects three representative periods of uptrend, downtrend, and fluctuation for backtesting analysis from the 14 divided intervals, as shown in Table \ref{tab:backtesting_intervals}.

\paragraph{\textbf{Model Backtesting Performance During Uptrend}}

\begin{figure}[!htbp]
	\centering
	\includegraphics[width=\columnwidth]{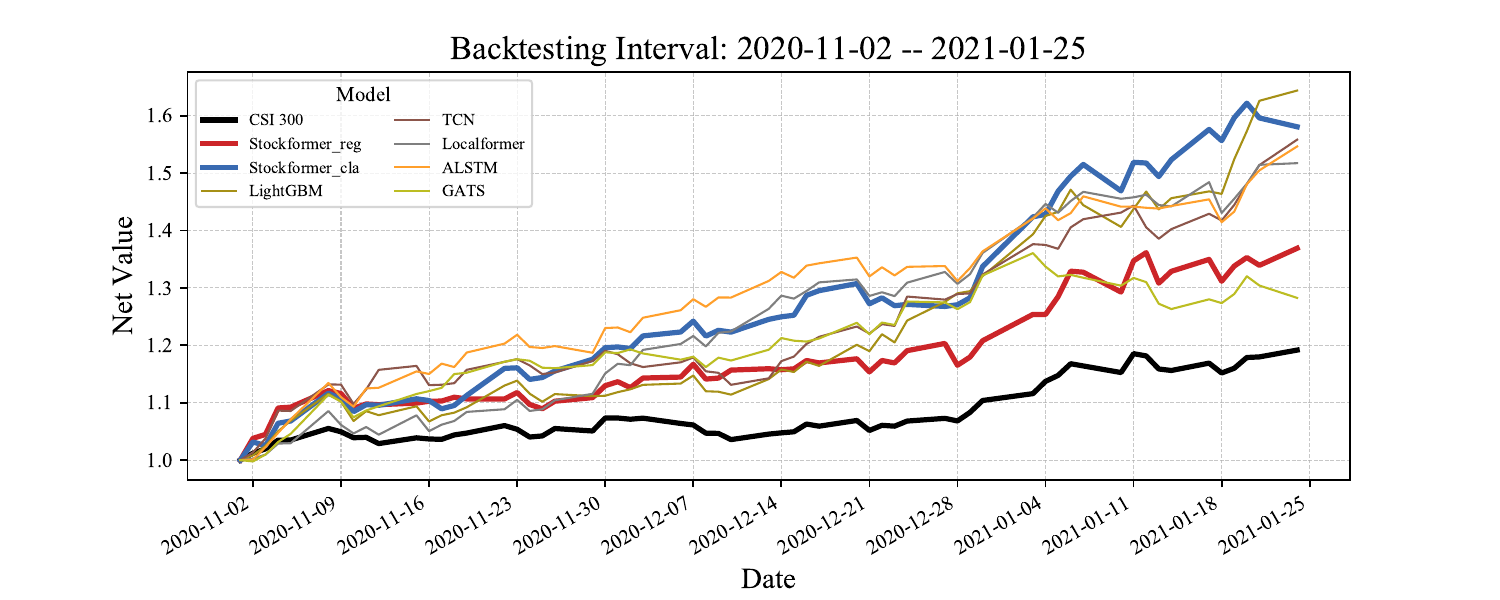} 
	\caption{Out-of-sample investment portfolio net value curves during the uptrend period (2020-11-02 to 2021-01-25). Bold lines represent the CSI 300 index benchmark (black), regression task output net value series (red), and classification task output net value series (blue). Thin lines represent selected outstanding baseline models (LightGBM, ALSTM, TCN, GATs, Localformer).} 
	\label{fig:Uptrend_Model_Performance} 
\end{figure}

During the backtesting of the uptrend market, this study selected the period from November 2, 2020, to January 25, 2021. The models used included Stockformer's regression output (Stockformer\_reg) and classification output (Stockformer\_cla), along with other selected high-performing baseline models. To determine which output type should be used during the backtesting period, we referred to the analysis of the high confidence prediction proportion indicator discussed in subsection \ref{subsec:backtest_input_selection}. The results indicate that using Stockformer's classification output (Stockformer\_cla) for predictions was more appropriate during this period. Furthermore, the CSI 300 index was used as a benchmark, as shown in Figure \ref{fig:Uptrend_Model_Performance}. All models' backtested net values were generally higher than the benchmark index. Particularly, the Stockformer model consistently outperformed other models in net value during most of the period, highlighting its superior performance and stability in uptrend markets.

\begin{table}[!htbp]
\centering
\caption{Investment Portfolio Performance during the Uptrend: \textbf{Bold} indicates best performance, \underline{underlined} indicates second best.  An upward arrow (\(\uparrow\)) indicates better performance with higher values, and a downward arrow (\(\downarrow\)) indicates better performance with lower values. This table lists various performance metrics for investment portfolios during the uptrend period in the test dataset (out-of-sample prediction) from November 2, 2020, to January 25, 2021.}
\label{tab:uptrend_performance}
\small 
\begin{tabular}{>{\raggedright\arraybackslash}p{3cm} >{\raggedright\arraybackslash}p{2.5cm} >{\raggedright\arraybackslash}p{2.5cm} >{\raggedright\arraybackslash}p{2.5cm} >{\raggedright\arraybackslash}p{2.5cm}}
\toprule
Benchmark \& Model & Annualized Return (\%)\(\uparrow\) & Annualized Volatility (\%)\(\downarrow\) & Maximum Drawdown (\%)\(\downarrow\) & Sharpe Ratio (\%)\(\uparrow\) \\
\midrule
CSI 300 & 80.52 & 16.46 & 3.5 & 4.73 \\
\midrule
LightGBM & \textbf{270.45} & 29.35 & 4.41 & \textbf{9.12} \\
ALSTM & 229.56 & \underline{26.82} & \underline{3.63} & 7.96 \\
TCN & 234.5 & 27.43 & 4.98 & 8.45 \\
GATs & 118.7 & \textbf{25.48} & 7.16 & 4.55 \\
Localformer & 217.31 & 26.91 & 3.79 & 7.98 \\
\midrule
\textbf{Stockformer\_cla} & \underline{239.73} & 29.78 & \textbf{3.07} & \underline{8.46} \\
\bottomrule
\end{tabular}
\end{table}

During the uptrend market period from November 2, 2020, to January 25, 2021, various prediction models demonstrated distinct performances as shown in Table \ref{tab:uptrend_performance}. LightGBM led with the highest annualized return of 270.45\%, indicating significant profitability during this period. ALSTM and TCN, with annualized returns of 229.56\% and 234.5\% respectively, also exhibited high profitability but were accompanied by relatively high volatility, which may increase investment risk. GATs showed good risk control capabilities with the lowest annual volatility of 25.48\%, although its returns were comparatively lower. Localformer achieved a better balance between risk and returns.

In contrast, the Stockformer model also performed impressively during this uptrend, achieving an annualized return of 239.73\%, second only to LightGBM. More importantly, Stockformer's maximum drawdown was only 3.07\%, significantly lower than other models, demonstrating excellent resilience during market downturns. Additionally, its Sharpe ratio of 8.46\% ranked second among all models, proving its efficiency in risk-adjusted returns. Overall, Stockformer not only excelled in profitability but also showcased significant advantages in risk management and market adaptability, making it an ideal choice in uptrend conditions.

\paragraph{\textbf{Model Backtesting Performance During Downtrend}}

\begin{figure}[!htbp]
	\centering
	\includegraphics[width=\columnwidth]{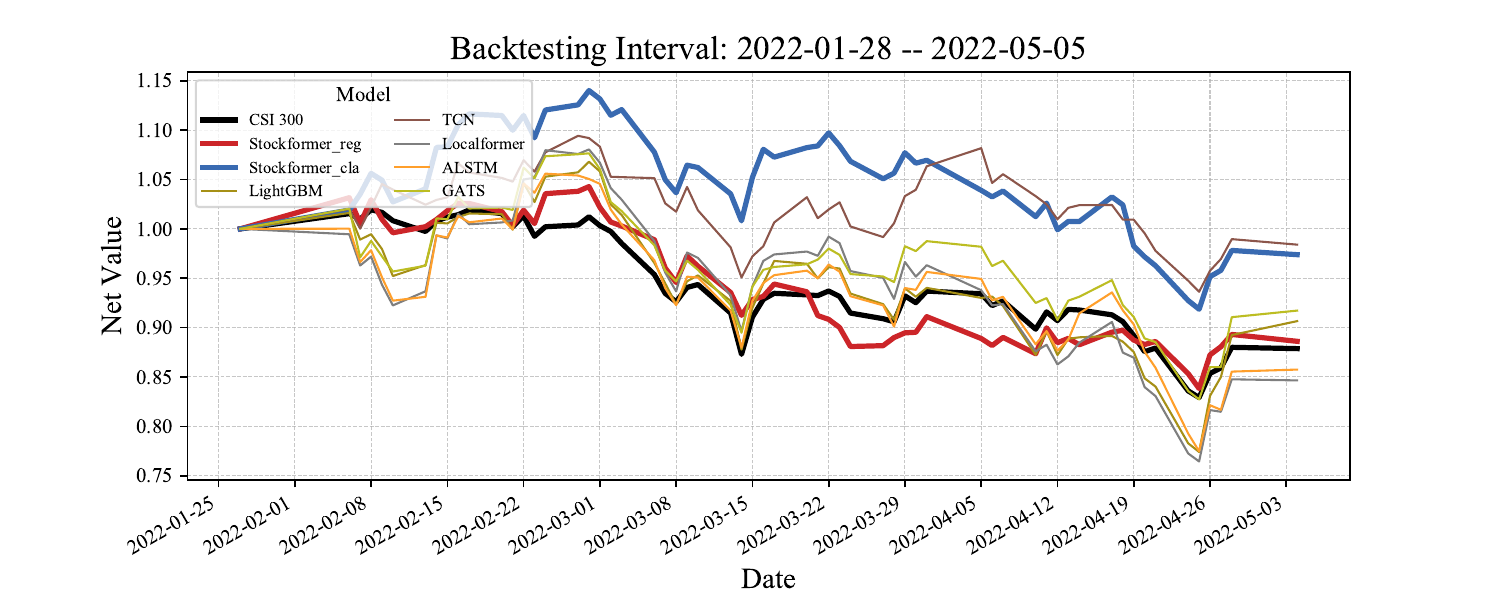} 
	\caption{Out-of-sample investment portfolio net value curves during the downtrend period (2022-01-28 to 2022-05-05). Bold lines represent the CSI 300 index benchmark (black), regression task output net value series (red), and classification task output net value series (blue). Thin lines represent selected outstanding baseline models (LightGBM, ALSTM, TCN, GATs, Localformer).} 
	\label{fig:downtrend_backtest_result} 
\end{figure}

\begin{table}[!htbp]
\centering
\caption{Investment Portfolio Performance During the Downtrend: \textbf{Bold} indicates best performance, \underline{underlined} indicates second best. An upward arrow (\(\uparrow\)) indicates better performance with higher values, and a downward arrow (\(\downarrow\)) indicates better performance with lower values. This table lists various performance metrics for investment portfolios during the downtrend period in the test dataset (out-of-sample prediction) from January 28, 2022, to May 5, 2022.}
\label{tab:downtrend_performance}
\small 
\begin{tabular}{>{\raggedright\arraybackslash}p{3cm} >{\raggedright\arraybackslash}p{2.5cm} >{\raggedright\arraybackslash}p{2.5cm} >{\raggedright\arraybackslash}p{2.5cm} >{\raggedright\arraybackslash}p{2.5cm}}
\toprule
Benchmark \& Model & Annualized Return (\%)\(\uparrow\) & Annualized Volatility (\%)\(\downarrow\) & Maximum Drawdown (\%)\(\downarrow\) & Sharpe Ratio (\%)\(\uparrow\) \\
\midrule
CSI 300 & -50.94 & 26.67 & 18.66 & -1.99 \\
\midrule
LightGBM & -39.25 & 42.61 & 27.52 & -0.97 \\
ALSTM & -59.85 & 43.62 & 26.63 & -1.42 \\
TCN & \textbf{-6.72} & \textbf{28.66} & \textbf{14.43} & \textbf{-0.31} \\
GATs & -34.77 & 38.11 & 23.15 & -0.97 \\
Localformer & -64.46 & 43.41 & 29.25 & -1.53 \\
\midrule
\textbf{Stockformer\_cla} & \underline{-15.18} & \underline{31.04} & \underline{19.60} & \underline{-0.56} \\
\bottomrule
\end{tabular}
\end{table}

During the analysis of backtesting performance in downtrend conditions, this study focused on the period from January 28, 2022, to May 5, 2022. For this interval, the Stockformer model's classification task output (Stockformer\_cla) and other distinguished baseline models were employed for comparison. All models exhibited a declining trend in their net value curves, reflecting the overall negative market trend. However, as shown in Figure \ref{fig:downtrend_backtest_result}, compared to other models and the CSI 300 index, Stockformer and TCN showed smaller declines in net value, demonstrating their relative resilience and stable excess returns during market downturns.

Table \ref{tab:downtrend_performance} presents the performance of various investment portfolios corresponding to each model during the downturn market conditions within the backtesting period (January 28, 2022, to May 5, 2022). Multiple stock prediction models displayed varying performances. LightGBM and GATs, despite showing lower decreases in annualized returns (-39.25\% and -34.77\%, respectively), faced higher annualized volatility and maximum drawdowns, indicating significant pressure under market fluctuations. ALSTM and Localformer exhibited even steeper declines and higher volatility, further highlighting the challenges in extreme market conditions.

In contrast, the Stockformer model demonstrated noticeable stability during this market downturn. It experienced a lesser decline in annualized returns (-15.18\%), ranking second best among all models, which illustrates its robust ability to resist market downturns. Additionally, Stockformer's maximum drawdown was 19.60\%, which, although not the lowest, still performed relatively well compared to other models. Regarding volatility, Stockformer also reported the second lowest annualized volatility (31.04\%), further confirming its robustness relative to other models.

\paragraph{\textbf{Model Backtesting Performance During Sideways Market}}

\begin{figure}[!htbp]
	\centering
	\includegraphics[width=\columnwidth]{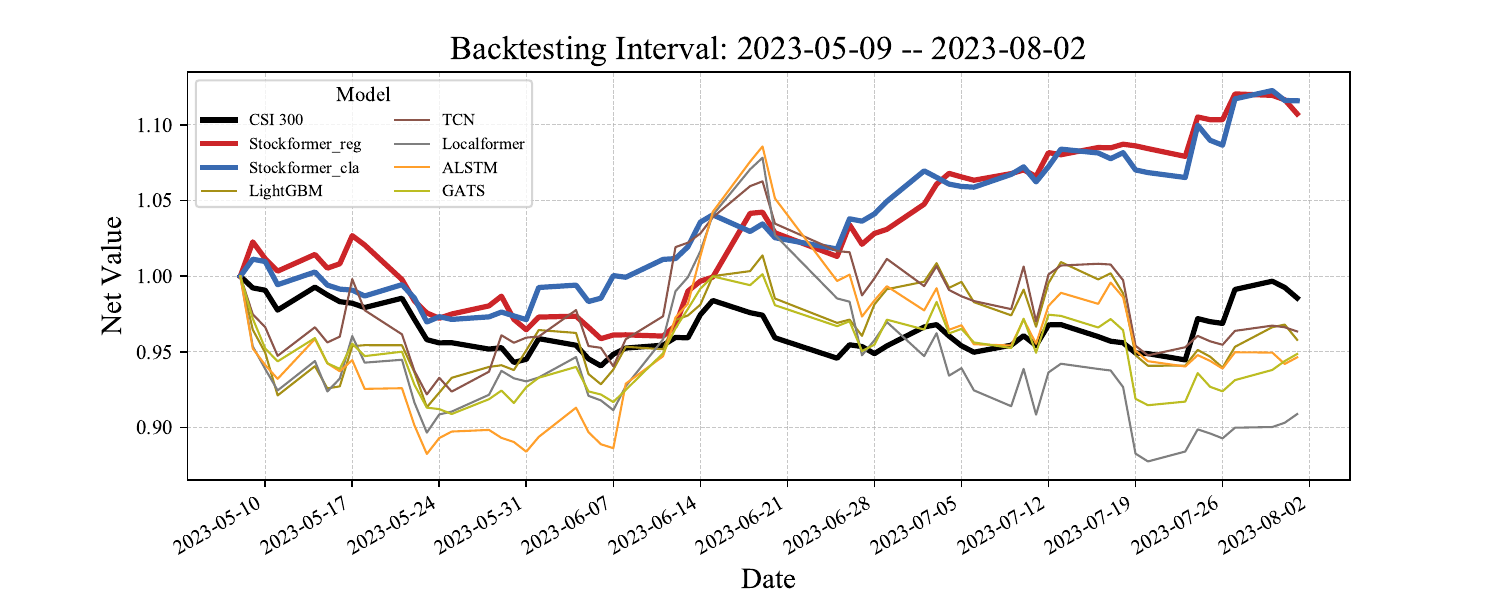} 
    \caption{Out-of-sample investment portfolio net value curves during the sideways market (2023-05-09 to 2023-08-02). Bold lines represent the CSI 300 index benchmark (black), regression task output net value series (red), and classification task output net value series (blue). Thin lines represent selected outstanding baseline models (LightGBM, ALSTM, TCN, GATs, Localformer).}
	\label{fig:sideways_market_backtest_result} 
\end{figure}

During the analysis of market performance in a sideways market condition, this study focused on the period from May 9, 2023, to August 2, 2023. For this interval, we utilized the classification output of the Stockformer model (Stockformer\_cla) along with other exemplary baseline models for comparison. As shown in Figure~\ref{fig:sideways_market_backtest_result}, during the sideways market conditions, Stockformer demonstrated a significant advantage. Its backtesting net value was positive, unlike most other models which predominantly showed negative net values. Furthermore, compared to the CSI 300 index, Stockformer was able to provide more stable excess returns, showcasing its exceptional adaptability and robustness in handling market fluctuations.

\begin{table}[htbp]
\centering
\caption{Investment Portfolio Performance During Sideways Market Conditions: \textbf{Bold} indicates the best performance, \underline{underlined} indicates the second best. An upward arrow (\(\uparrow\)) indicates better performance with higher values, and a downward arrow (\(\downarrow\)) indicates better performance with lower values. This table lists various performance metrics for investment portfolios during the sideways market conditions in the test dataset (out-of-sample prediction) form May 9, 2023, to August 2, 2023.}
\label{tab:sideways_performance}
\small 
\begin{tabular}{>{\raggedright\arraybackslash}p{3cm} >{\raggedright\arraybackslash}p{2.5cm} >{\raggedright\arraybackslash}p{2.5cm} >{\raggedright\arraybackslash}p{2.5cm} >{\raggedright\arraybackslash}p{2.5cm}}
\toprule
Benchmark \& Model & Annualized Return (\%)\(\uparrow\) & Annualized Volatility (\%)\(\downarrow\) & Maximum Drawdown (\%)\(\downarrow\) & Sharpe Ratio (\%)\(\uparrow\) \\
\midrule
CSI 300 & -6.05 & 14.30 & 5.92 & -0.56 \\
\midrule
LightGBM & -17.64 & 25.28 & \underline{8.65} & -0.78 \\
ALSTM & -22.51 & 30.94 & 13.52 & -0.79 \\
TCN & \underline{-15.31} & 28.26 & 10.82 & \underline{-0.61} \\
GATs & -21.57 & \underline{23.09} & 9.10 & -1.02 \\
Localformer & -38.25 & 32.86 & 18.61 & -1.22 \\
\midrule
\textbf{Stockformer\_cla} & \textbf{44.48} & \textbf{15.68} & \textbf{4.14} & \textbf{2.71} \\
\bottomrule
\end{tabular}
\end{table}

The performance of investment portfolios during sideways market conditions is presented in Table~\ref{tab:sideways_performance}. During this phase, most models failed to achieve positive returns, reflecting the challenging nature of sideways markets on model stability. LightGBM, ALSTM, TCN, GATs, and Localformer all exhibited high annualized volatility and maximum drawdowns, with ALSTM and Localformer performing the worst, experiencing annualized returns dropping to -22.51\% and -38.25\% respectively. This indicates extreme sensitivity in sideways markets, leading to significant losses.

In contrast, Stockformer significantly outperformed other models under these market conditions, with an annualized return of 44.48\%, and maintained lower annualized volatility (15.68\%) and the smallest maximum drawdown (4.14\%). Its Sharpe ratio reached 2.71, far surpassing all other models. This demonstrates that Stockformer not only managed to maintain positive returns during market fluctuations but also effectively controlled risk and preserved high return stability. This highlights Stockformer's superior performance in managing market uncertainties and its high risk-adjusted returns, making it an ideal choice for investing in sideways markets.

\paragraph{\textbf{Summary Analysis of Backtesting Strategy}}

\begin{figure}[!htbp]
    \centering
    \includegraphics[width=\textwidth]{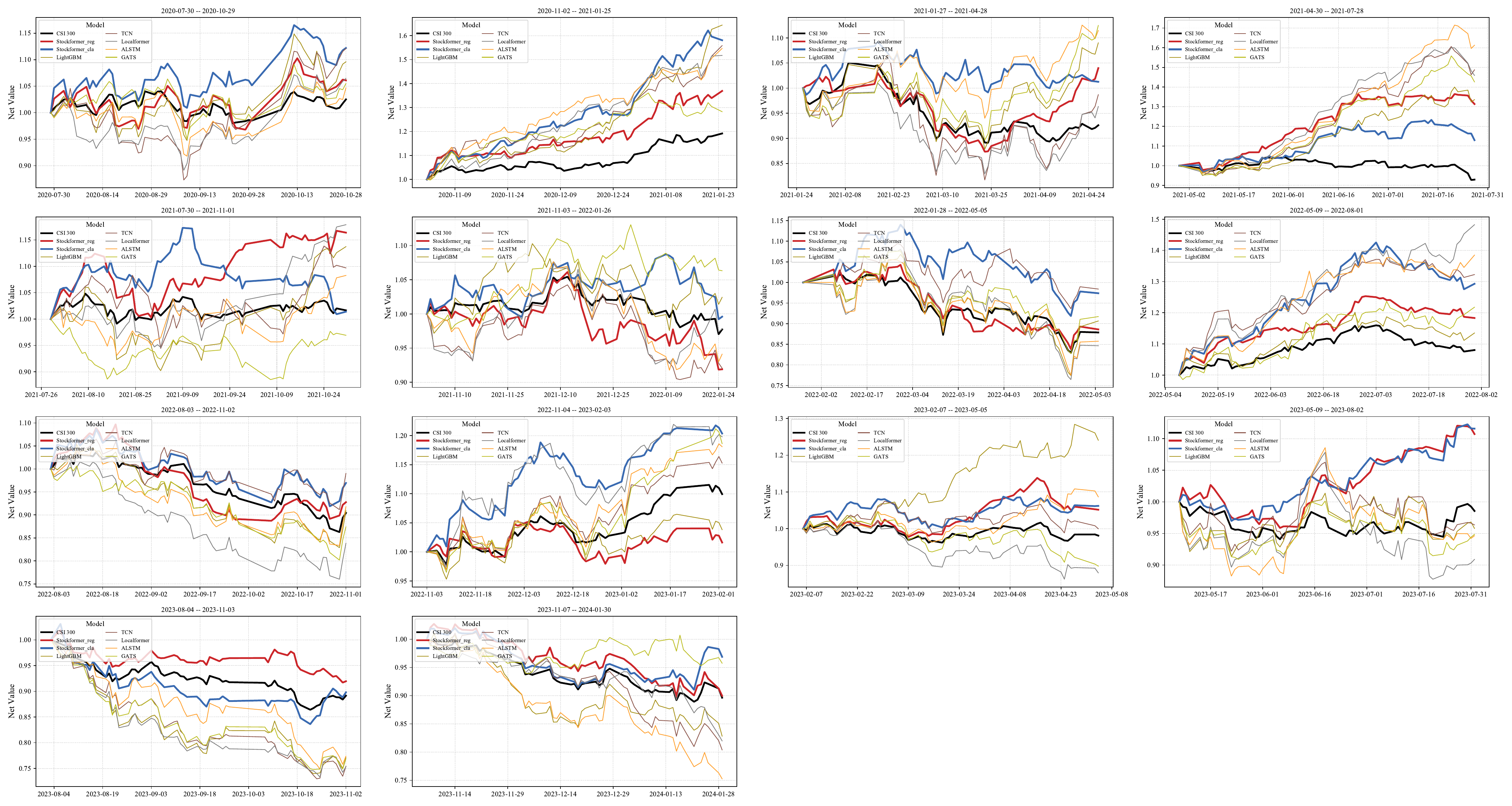}
    \caption{Comprehensive chart of net value changes for investment portfolios in various time intervals based on out-of-sample predictions. Bold lines represent the CSI 300 index benchmark (black), Stockformer's regression task output net value series (red), and classification task output net value series (blue). Thin lines display the net value series for selected outstanding baseline models including LightGBM, ALSTM, TCN, GATs, and Localformer.}
    \label{fig:backtest_results}
\end{figure}

From the backtesting analysis across uptrend, downtrend, and sideways market conditions, it is evident that Stockformer demonstrates significant advantages over other models. In uptrend markets, Stockformer showed strong profitability, closely matching LightGBM, but exhibited superior performance in terms of risk control and maximum drawdown. In downtrend markets, although all models experienced declines in net value, Stockformer's losses were noticeably smaller than those of other models, indicating higher market adaptability and stability. In sideways markets, Stockformer was the only model to achieve positive returns, and its risk-adjusted returns (Sharpe ratio) far exceeded those of other models, emphasizing its ability to maintain consistent profitability in complex market environments.

\begin{table}[htbp]
\centering
\caption{Comparison of investment strategy performance over multiple periods with the CSI 300 Index: Analysis of annualized return rates, Sharpe ratios, and maximum drawdowns}
\label{tab:backtesting_performance}
\small
\begin{tabular}{c|cc|cc|cc}
\hline
Backtesting Period & \multicolumn{2}{c|}{Annualized Return (\%)\(\uparrow\)} & \multicolumn{2}{c|}{Maximum Drawdown (\%)\(\downarrow\)} & \multicolumn{2}{c}{Sharpe Ratio (\%)\(\uparrow\)} \\
\cline{2-7}
 & Stockformer & CSI 300 & Stockformer & CSI 300 & Stockformer & CSI 300 \\
\hline
2020-08 -- 2020-10 & 0.4703 & 0.1053 & 0.0772 & 0.0580 & 1.5611 & 0.4611 \\
2020-11 -- 2021-01 & 2.3973 & 0.8052 & 0.0307 & 0.0350 & 7.9580 & 4.7254 \\
2021-02 -- 2021-04 & 0.0107 & -0.3106 & 0.0953 & 0.1518 & -0.0611 & -1.4306 \\
2021-05 -- 2021-07 & 1.2751 & -0.2976 & 0.0438 & 0.1105 & 5.0450 & -1.7625 \\
2021-08 -- 2021-10 & 0.6465 & 0.0694 & 0.1087 & 0.0543 & 2.4503 & 0.2964 \\
2021-11 -- 2022-01 & -0.0588 & -0.0948 & 0.0882 & 0.0797 & -0.4133 & -0.8723 \\
2022-02 -- 2022-05 & -0.1518 & -0.5094 & 0.1960 & 0.1866 & -0.5566 & -1.9882 \\
2022-05 -- 2022-08 & 1.1880 & 0.3371 & 0.1061 & 0.0725 & 4.0540 & 1.9663 \\
2022-08 -- 2022-11 & -0.1703 & -0.4019 & 0.1559 & 0.1680 & -0.7093 & -2.2679 \\
2022-11 -- 2023-02 & 0.8131 & 0.4175 & 0.0688 & 0.0425 & 3.7037 & 2.6248 \\
2023-02 -- 2023-05 & 0.1748 & -0.0794 & 0.0783 & 0.0497 & 0.9286 & -0.7705 \\
2023-05 -- 2023-08 & 0.4448 & -0.0605 & 0.0414 & 0.0592 & 2.7118 & -0.5607 \\
2023-08 -- 2023-11 & -0.3799 & -0.4559 & 0.1011 & 0.1359 & -2.6577 & -3.556 \\
2023-11 -- 2024-01 & -0.4633 & -0.4348 & 0.1268 & 0.1107 & -2.5537 & -3.2945 \\
\hline
\end{tabular}
\end{table}

From July 2020 to January 2024, the investment strategy proposed in this paper was meticulously backtested across 14 distinct time intervals in sub-datasets. The net value curves of these models are presented in Figure \ref{fig:backtest_results}, while Table \ref{tab:backtesting_performance} provides detailed backtesting performance across these intervals. The results indicate that, except for the period from November 2023 to January 2024, the strategy's annualized return rate outperformed the CSI 300 index in the remaining 13 intervals. Moreover, the Sharpe ratio exceeded that of the CSI 300 index in all 14 intervals, and the maximum drawdown was similar to that of the CSI 300, demonstrating the strategy's effective risk control. From the various sub-figures, it is observable that the Stockformer model secured stable excess returns in all test intervals and, in most cases, its portfolio net value performance surpassed other selected outstanding baseline models such as LightGBM, ALSTM, etc. In summary, the investment strategy described in this paper demonstrates exceptional stability and considerable excess returns relative to the CSI 300 index across varying market conditions, whether in rising, falling, or volatile markets.

These results demonstrate that the Stockformer model exhibits outstanding stability and reliability across diverse market conditions. Particularly during market downturns or volatile periods, it maintains high performance levels, showing a high degree of adaptability to market fluctuations. This capability allows Stockformer to be suitable not only for strategies aiming for high returns but also excels in environments requiring stringent risk control. Hence, whether the investment goal is capital appreciation or capital preservation, Stockformer has proven to be a robust investment tool capable of providing solid support under various market conditions. Furthermore, for investors seeking consistent performance across different market environments, Stockformer offers an effective strategy option. Through detailed backtesting verification, Stockformer has demonstrated its potential and practicality as an advanced investment strategy.


\section{Conclusion and Future Work}

\subsection{Conclusion}
This study introduces Stockformer, a stock selection model that integrates wavelet transform and multitask self-attention networks, aiming to enhance the precision and adaptability of analyzing the complex dynamics of global securities markets. By extensively mining data from the stocks within the CSI 300 index and partitioning it into 14 subdatasets, and by employing 360 rigorously tested price-volume factors, the model significantly improves the interpretation of market fluctuations and abrupt events. Stockformer utilizes advanced deep learning technologies, including dual-frequency spatiotemporal encoder, graph embedding techniques, and multitask learning strategies. These technologies not only increase the accuracy of stock return and market trend predictions but also enhance the model’s capability to detect market dynamics in detail. Specifically, the model precisely captures short-term market fluctuations and long-term trends through high-low frequency decomposition techniques, while the integrated dual-frequency spatiotemporal encoder ensures effective processing of temporal and spatial dependencies. Experimental results demonstrate that Stockformer outperforms ten baseline models across various predictive performance metrics. Notably, it achieves a directional accuracy of 57.46\%, significantly higher than other models. Moreover, the study conducts backtesting on different stock selection models, assessing their performance in bullish, bearish, and volatile markets. Backtesting results indicate that the Stockformer-based stock selection model achieves annualized returns of 239.73\%, -15.18\%, and 44.48\% in rising, falling, and fluctuating markets, respectively; with maximum drawdowns of 3.07\%, 19.6\%, and 4.14\%; and Sharpe ratios of 8.46, -0.56, and 2.71. Compared to other models, Stockformer displays more stable performance across different market conditions and secures stable excess returns over the CSI 300 index.

In conclusion, Stockformer demonstrates the potential of deep learning technologies in analyzing complex markets and provides financial market analysts with a valuable decision-support tool, aiding them in making wiser investment decisions in volatile market environments.

\subsection{Future Work}

Despite the significant achievements of the Stockformer model in the application to stock markets, future research and development are still filled with challenges and opportunities. Firstly, the wavelet transform used in the model involves multiple complex parameter settings, such as periodic scales and transformation depths. The determination of these parameters is both time-consuming and requires repetitive experimentation. To streamline this process, future work will explore how to automatically identify and optimize these periodic parameters within the neural network architecture, thereby alleviating the burden of preliminary data processing.

Secondly, the current model is based on a static pool of CSI 300 stocks. However, the dynamic nature of the stock pool is a fundamental characteristic of stock markets, and using a fixed stock pool might limit the model's applicability and predictive accuracy. Therefore, there are plans to develop a dynamic updating mechanism that allows Stockformer to adapt to changes in the stock pool, thus enhancing the precision of its predictions.

Additionally, as new stocks are continuously introduced, the model needs to be able to adapt quickly to these changes. To this end, we plan to employ meta-learning approaches to enhance the adaptability of the model, allowing it to maintain efficient predictive capabilities under unknown market conditions. With these improvements, Stockformer will be better equipped to serve financial market analysts, helping them make more precise decisions in a volatile market environment.

\section*{Acknowledgements}

This research is supported by the National Natural Science Foundation of China under grant no. 12001556, the National Key Research and Development Program of China under the ``National Key R\&D Program of China (no. 2023YFF0614700)", the Program for Innovation Research in Central University of Finance and Economics, and the Central University of Finance and Economics Postgraduate Thesis Competition.

\bibliography{sample}

\end{document}